\documentclass[]{aa}
\usepackage{graphicx} 
\usepackage[pdfencoding=auto,psdextra]{hyperref}
\usepackage{natbib}
\usepackage{amsmath}	
\usepackage{amssymb}	
\usepackage{txfonts}
\usepackage{xcolor}

\newcommand*{\neii}{{[$\ion{\rm Ne}{II}$]13$\mu$m\,}}
\newcommand*{\oiv}{{[$\ion{\rm O}{IV}$]26$\mu$m\,}}
\newcommand*{\feii}{{[$\ion{\rm Fe}{II}$]26$\mu$m\,}}
\newcommand*{\niii}{{[$\ion{\rm N}{III}$]57$\mu$m\,}}
\newcommand*{\oiiia}{{[$\ion{\rm O}{III}$]52$\mu$m\,}}
\newcommand*{\oiiib}{{[$\ion{\rm O}{III}$]88$\mu$m\,}}
\newcommand*{\spr}{{\sc Spritz}\xspace}
\newcommand*{\hers}{\textit{Herschel}\xspace}
\newcommand*{\jwst}{JWST\xspace}

\nolinenumbers

\title{Disentangling the co-evolution of galaxies and supermassive black holes with PRIMA} 

\author{Bisigello, L.$^{1,2}$\thanks{\email{laura.bisigello@inaf.it}}, Gruppioni$^{3}$, C., Bolatto A.$^{4}$, Ciesla, L.$^{5}$, Pope, A.$^{6}$, Armus, L.$^{7}$ , Smith, J.D.$^{8}$, Somerville, R. S.$^{9}$, Yung, L. Y. A.$^{10}$, Wright, R. J.$^{11}$, Bradford, C. M.$^{13,14}$, Glenn, J.$^{12}$, and Feltre, A.$^{15}$}
\authorrunning{Bisigello, L. et al.}
\titlerunning{Disentangling galaxy and AGN co-evolution using far-IR spectro-photometric observations}
\institute{$^{1}$ INAF, Istituto di Radioastronomia, Via Piero Gobetti 101, 40129 Bologna, Italy\\
$^{2}$ Dipartimento di Fisica e Astronomia "G. Galilei", Universit\`a di Padova, Via Marzolo 8, 35131 Padova, Italy \\
$^{3}$ INAF-Osservatorio di Astrofisica e Scienza dello Spazio, via Gobetti 93/3, I-40129, Bologna, Italy\\
$^{4}$Department of Astronomy, University of Maryland, College Park, MD 20742, USA\\
$^{5}$Aix Marseille Univ, CNRS, CNES, LAM, Marseille, France\\
$^{6}$Department of Astronomy, University of Massachusetts, Amherst, MA 01003, USA \\
$^{7}$ IPAC, California Institute of Technology, 1200 E. California Boulevard, Pasadena, CA 91125, USA \\
$^{8}$Ritter Astrophysical Research Center, University of Toledo, Toledo, OH 43606, USA\\
$^{9}$Center for Computational Astrophysics, Flatiron Institute, 162 5th Ave, New York, NY 10010, USA \\
$^{10}$Space Telescope Science Institute, 3700 San Martin Dr., Baltimore, MD 21218, USA\\
$^{11}$Department of Physics, University of Helsinki, Gustaf Hällströmin katu 2, FI-00014 Helsinki, Finland\\
$^{12}$NASA Goddard Space Flight Center, Greenbelt, MD 20771, USA\\
$^{13}$NASA Jet Propulsion Laboratory, 4800 Oak Grove Dr, Pasadena, CA 91011, USA\\
$^{14}$California Institute of Technology, 1200 E California Blvd, Pasadena, CA 91125, USA\\
$^{15}$ INAF-Osservatorio Astrofisico di Arcetri, Largo E. Fermi 5, I-50125, Firenze, Italy}
\date{}

\begin{document}

\abstract{
    The most active phases of star formation and black hole accretion are strongly affected by dust extinction, making far-infrared (far-IR) observations the best way to disentangle and study the co-evolution of galaxies and super massive black holes. The plethora of fine structure lines and emission features from dust, ionised and neutral atomic and warm molecular gas in the rest-frame mid- and far-IR provide unmatched diagnostic power to determine the properties of gas and dust, measure gas-phase metallicities and map cold galactic outflows in even the most obscured galaxies. By combining multi-band photometric surveys with low and high-resolution far-IR spectroscopy, the PRobe far-Infrared Mission for Astrophysics (PRIMA), a concept for a far-IR, 1.8m-diameter, cryogenically cooled observatory, will revolutionise the field of galaxy evolution by taking advantage of this IR toolkit to find and study dusty galaxies across galactic time. In this work, we make use of the phenomenological simulation \spr{} and the Santa Cruz semi-analytical model to describe how a moderately deep multi-band PRIMA photometric survey can easily reach beyond previous IR missions to detect and study galaxies down to $10^{11}\,L_{\odot}$ beyond cosmic noon and at least up to $z=4$, even in the absence of gravitational lensing. By decomposing the spectral energy distribution (SED) of these photometrically selected galaxies, we show that PRIMA can be used to accurately measure the relative AGN power, the mass fraction contributed by polycyclic aromatic hydrocarbon (PAH) and the total IR luminosity. At the same time, spectroscopic follow up with PRIMA will allow to trace both the star formation and black hole accretion rates (SFR, BHAR), the gas phase metallicities and the mass outflow rates of cold gas in hundreds to thousands of individual galaxies to $z=2$. 
    }

\maketitle
\nolinenumbers

\section{Introduction}

Super massive black holes (SMBHs) of millions of solar masses seem to be ubiquitous at the centre of galaxies in the local Universe \citep[e.g.][]{Magorrian1998,Gultekin2009,McConnell2011}. Observations have pointed out the presence of relations between the mass of the SMBH and many properties of the host galaxy, such as the stellar mass and the velocity dispersion of the galaxy bulge \citep{Magorrian1998,Ferrarese2002}, the halo mass \citep[e.g.][]{Ferrarese2002} and the stellar mass of the host itself \citep[e.g.][]{Mullaney2012,Reines2015}. The presence of these scaling relations suggests a close evolutionary link between the SMBH and its host galaxy \citep{Kormendy2013}. \par
Co-evolution between SMBHs and their host galaxies is also supported by the similarity between the time evolution of the cosmic star-formation-rate density \citep[SFRD; e.g.][]{Gruppioni2013,Madau2014,Traina2024} and of the black-hole accretion rate density \citep[BHARD; e.g.][]{Delvecchio2014}. Indeed, both quantities show an increase from early epochs up to $z\sim2-3$, followed by a rapid decrease, although these measures are made from disjoint samples. While the rough shape of the SFRD with cosmic time can be understood in the context of merger driven evolution, theoretical models calibrated to reproduce present day scaling relations, seem to produce a wide range in the BHARD at higher redshifts \citep{Habouzit2020,Habouzit2021} due to differences in the implementation of supernovae and black-hole feedback and sub-grid physics.\par
Observationally, the study of both the SFRD and the BHARD is hampered by the presence of dust, as the most active phases of black hole growth and star formation are heavily enshrouded. Indeed, the contribution of the dust-obscured SFRD to the total is above 50\% at least up to $z=4$ \citep[][Traina et al. in prep.]{Zavala2021}. Moreover, some galaxies are so obscured by dust to be detectable in the infrared (IR), but faint or totally missed at optical wavelengths \citep[e.g.][]{Gruppioni2020,Talia2021,Rodighiero2023,Bisigello2023b,PerezGonzalez2023}. \par
\hers results showing that the majority of IR galaxies with $\rm L_{IR}>10^{11}\,L_{\odot}$ at $z>1$ host a low-luminosity or obscured AGN \citep[AGN;][]{Gruppioni2013,Magnelli2013} support the hypothesis of SMBHs and galaxies co-growth and co-evolution, and therefore the need to observe the peak of their activity (i.e., cosmic noon) in the IR. Similarly, recent \textit{James Webb Space Telescope} (\jwst) observations of numerous faint, broad-line AGN at $z>5$ \citep[e.g.][]{onoue23,Kocevski23b, Harikane23, Larson23}, possibly pointing to Eddington-limited accretion onto massive black hole seeds formed at $z>15$ \citep[e.g.,][]{Maiolino2023}. At the same time, about 20\% of the broad-line AGN identified with \jwst appear as red and compact sources, with a steep red continuum in the rest-frame optical and a blue/flat continuum in the rest-frame ultraviolet (UV) \citep[e.g.][]{Kocevski23b,Kocevski2024,Harikane23,Matthee2023,Greene23,Killi23}. Despite the AGN signature in the rest-frame optical, the majority of  these so called `little red dots' are weak or undetected X-ray emission with Chandra \citep{Yue2024}. These objects could be heavily obscured by dust heated by buried AGN \citep[e.g.][]{Kocevski23b,Matthee2023,Greene23}, very compact starbursts with large numbers of OB stars \citep[e.g.,][]{PerezGonzalez2024}, Compton thick AGN with a dust-poor medium or intrinsically X-ray weak AGN, such as Narrow Line Seyfert 1s \citep{Maiolino2024}.
\par

\begin{figure}
    \centering
    \includegraphics[trim=5 3 5 5,clip,width=\columnwidth,keepaspectratio]{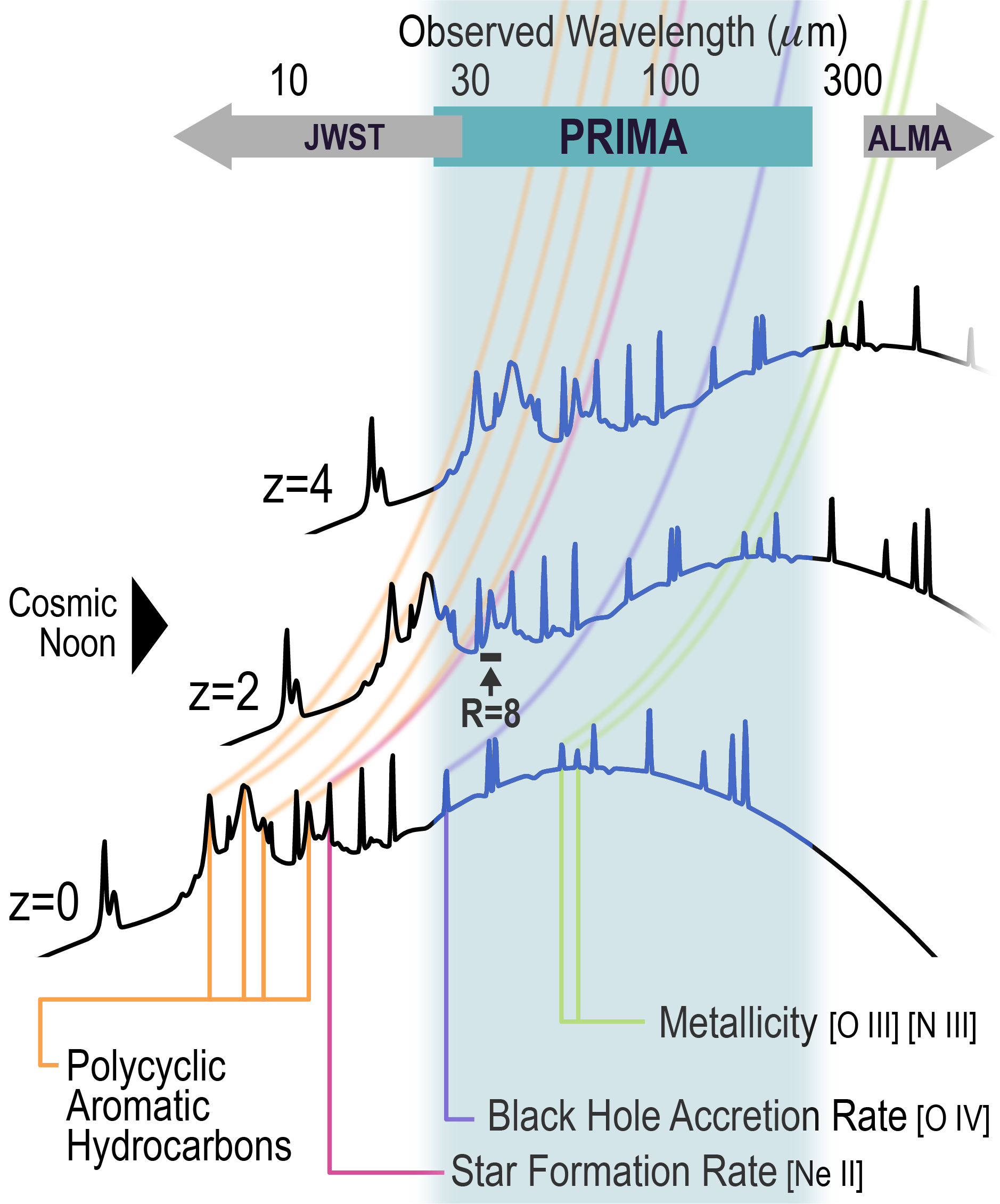}
    \caption{The mid- and far-IR spectra of galaxies contains a plethora of features useful to derive SFR, BHAR and metallicity. The observed wavelengths of some of these tracers are reported in the figure, together with the SED of an example star-forming galaxy at $z=0$, $z=2$ and $z=4$, taken from \citet[][section 47]{PRIMAGO} We also highlight the wavelength coverage of \jwst, PRIMA (shaded region) and ALMA.}
    \label{fig:irlines}
\end{figure}

Given the observational and theoretical evidence that galaxies and their SMBH grow together, and that the most rapid growth occurs during phases that are hidden by dust \citep{Hickox2018}, it is essential to measure SFR and BHAR in high-z populations using tools that are insensitive to the effects of dust obscuration.  It is also important to perform these studies in the same populations of galaxies, where observation biases can be well understood and selection effects minimised. The rest frame mid- and far-IR spectrum offers a plethora of fine structure gas lines and emission features from dust and warm molecular gas that are sensitive to AGN and star-formation (SF) activity (Fig. \ref{fig:irlines}) and are not (or very little) affected by dust-extinction. In particular, among the AGN tracers there are [$\ion{\rm Ne}{V}$] at 14.3 and 24.3 $\mu m$,  [$\ion{\rm O}{IV}$] at 24.9 $\mu m$ and [$\ion{\rm Ne}{III}$] at 15.7 $\mu$m \citep[e.g][]{Rigby2009,FernandezOntiveros2016,Feltre2023}. At the same time, Polycyclic Aromatic Hydrocarbon (PAH) emission features present between 3 and 20 $\mu m$, the 
mid-IR pure rotational lines of H2 at 17$\mu m$, low-ionisation fine structure lines like the [$\ion{\rm Ne}{II}$] line at 12.8 $\mu m$, the [$\ion{\rm S}{III}$] lines at 18 and 33 $\mu m$, or the [$\ion{\rm C}{II}$] line at 158 $\mu m$ \citep[e.g.][]{DeLooze2014,Vallini2015} are sensitive tracers of star formation. While some of these lines are excited by both star formation and AGN activity, it has been shown that the relative contributions can be decomposed \citep{Stone2022}. Many of these lines have already been used in the Local Universe to study the SFR and the BHAR in individual bright IR galaxies \citep[e.g.][]{Armus2004,Armus2006,Armus2007,Armus2023,Spinoglio2015,Spinoglio2022,Spoon2009,Diaz-Santos2013,Inami2013,Stierwalt2014,Lai2022,Stone2022}, but cold sensitive far-IR observatories are necessary to extend these studies to higher redshifts and to less extreme galaxies. \par

Dust grains play a crucial role on the formation and cooling of molecular gas and, therefore, on the formation of new stars \citep{Draine2003,Peimbert2010}. For this reason, theoretical models predict that the dust abundance and metallicity of the interstellar medium (ISM) should be closely related \citep{Zhukovska2014}, as a large fraction of metals is converted into dust when they are injected into the ISM by asymptotic giant branch stars and supernovae. At the same time, the rate of dust destruction by supernova blast waves is thought to be an order of magnitude higher than dust production by dying stars \citep{Dwek2011}. There is clearly a large gap in our knowledge of dust formation (and star-formation as a consequence) and destruction, which can be solved only by tracing both metals and dust in the same galaxies. PAHs features, present in the rest-frame mid-IR, are both tracers of star-formation and sensitive tracers of the abundance of small carbonaceous dust grains. At the same time, in dusty galaxies gas metallicity can be measured using bright, far-IR tracers  such as \niii, \oiiia and \oiiib \citep[][]{Nagao2011,Pereira-Santaella2017,FernandezOntiveros2017}, which are  unaffected by dust extinction and have a negligible dependence on the gas temperature, unlike their optical and UV counterparts \citep[e.g.,][]{BernardSalas2001}. \par

To fully understand the interplay between AGN and their host galaxies, it is also necessary to identify and measure galactic outflows, a powerful form of feedback that can inject significant amounts of energy into the ISM, and drive gas and dust into the circumgalactic medium (CGM). Outflows are invariably multi-phase, but the cold ($T<10^{4} K$) component often dominates the gas mass  \citep{Veilleux2020}. Spectroscopic observations of strong absorption features in the far-IR, such as OH, can be used to estimate the outflow velocity and mass outflow rate \citep{Gonzalez-Alfonso2014,Gonzalez-Alfonso2017}. Observations of cold outflows using OH absorption lines have been very effective, indicating large masses of outflowing dense gas that can rival or exceed the star formation rate in ultra-luminous IR galaxies \citep{Gonzalez-Alfonso2017}, although detailed studies have so far been limited to small numbers of galaxies in the Local Universe \citep[see][for a review]{Veilleux2020}. Sensitive, cold far-IR spectroscopic observatories that can reach levels significantly fainter than were possible with \hers{} are necessary to carry these studies to high-redshifts and trace cold outflows to cosmic noon and beyond in statistically significant samples of normal galaxies. 
\par

The PRobe far-Infrared Mission for Astrophysics, PRIMA\footnote{\href{https://prima.ipac.caltech.edu/}{https://prima.ipac.caltech.edu/}} (P.I. J. Glenn), is a concept for a far-IR, cryogenically cooled observatory with an 1.8 m diameter. PRIMA's current design includes two science instruments: FIRESS and PRIMAger. FIRESS is a powerful, multi-mode survey spectrometer covering wavelengths between 24 and 235 $\mu$m and having two different spectral modes, a low-resolution mode with $R\sim100$ and a high-resolution FTS mode which offers tunable spectral resolution up to $R=$4,400 at 112$\,\mu m$ and 20,000 at 25 $\,\mu$m. PRIMAger is a multiband spectro-photometric imager that offers hyper-spectral linear variable filters in two bands  ($R=10$,PHI1 and PHI2) from 24 to 80 $\mu$m together with polarimetric capabilities in four broadband filters (PPI) between 80 and 261 $\mu$m. Because it is cooled, and uses state of the art kinetic inductance detectors \citep[e.g.,][]{Day2003,Day2024,Baselmans2012}, PRIMA is orders of magnitude more sensitive than previous far-IR space missions. PRIMA is designed to rapidly survey the IR sky, achieving mapping speeds 3-5 orders of magnitude faster than its far-IR predecessors. In this work we outline the revolutionary role that PRIMA will play in studying galaxy and AGN co-evolution across cosmic time. Although here we illustrate the power of PRIMA by describing possible deep and wide surveys and follow-up spectroscopy with PRIMAger and FIRESS, we must stress that PRIMA is a true community observatory, with at least $75\%$ of the observing time available to astronomers through a traditional, peer-reviewed, General Observers program \citep[for an overview of different general observers science cases see][]{PRIMAGO}. Moreover, all PI data will be publicly available to Guest Investigators.

The paper is organised as follows. In Sect. \ref{sec:methods} we present the PRIMA example surveys together with the simulations used in the analysis. In Sect. \ref{sec:results} we discuss the SFR, BHAR, metallicity and outflows measurements, using both photometric and spectroscopic IR observations. We finally report and summarize our conclusions in Sect. \ref{sec:summary}. Throughout the paper, we consider a $\Lambda$CDM cosmology with $H_0=70\,{\rm km}\,{\rm s}^{-1}\,{\rm Mpc}^{-1} $, $\Omega_{\rm m}=0.27$, $\Omega_\Lambda=0.73$, and a Chabrier initial mass function  \citep[IMF,][]{Chabrier2003}.

\section{Methods}\label{sec:methods}
To assess the capability of PRIMA to disentangle galaxy and AGN evolution as well as to measure metallicity and cold outflows, we have made use of the spectro-photometric realisations of infrared-selected targets at all-$z$ \citep[\spr{}\footnote{\href{http://spritz.oas.inaf.it/}{http://spritz.oas.inaf.it/}} v1.13;][]{Bisigello2021,Bisigello2022} and the Santa Cruz semi-analytic model \citep[SC SAM;][]{Somerville2015}. In the next sections we give some brief details on the PRIMAger instrumental capabilities, on these two simulations, and explain how it is possible to measure both SFR and BHAR using far-IR spectro-photometric data.

\subsection{PRIMA surveys}\label{sec:prima}

As example cases, we considered two different designs for the PRIMAger surveys. Both surveys have a total observing time of 1500h, distributed over 1 deg$^{2}$ for the Deep survey and over 10 deg$^{2}$ for the Wide survey. In the rest of the paper, we consider a galaxy as detected if it has a $S/N>5$ in half of the PRIMAger hyper-spectra filters (PHI), to assure good description of the overall far-IR spectral energy distribution (SED) and limiting problems due to confusion. In Tab. \ref{tab:depths} we list the PRIMAger filters and the depths considered for the Deep and Wide surveys. As reference we use the conservative depths, taking into account that the instrument current best estimate (CBE) is better than this conservative expectations by a factor of $\sim1.7$ in the hyper-spectra PHI bands and $\sim2$ in the long wavelengths PPI filters. However, when necessary, we also make comparison with the payload limits, which are the guarantee requirement payload depths.
Although the hyper-spectra filters PHI band of PRIMAger will provide continuous, highly over-sampled spectra at $R = 10$ from 24 to 80$\,\mu m$, we assume in the following that this band is composed of 12 top-hat filters with an $R=10$. Similarly, we represent the four PRIMAger broad band filters (PPI) as continuous rectangular filters spanning the wavelength range of the band, i.e. $R=4$.

\begin{table}[]
    \caption{PRIMAger filter bandpasses and survey depths.} 
    \label{tab:depths}
    \centering
    \begin{tabular}{cc|cc|cc}
        & & \multicolumn{2}{|c|}{Conservative} & \multicolumn{2}{|c}{Payload} \\
        & & Deep & Wide & Deep & Wide \\
        Filter & $\lambda_{cen}$ & $f_{5\sigma}$ & $f_{5\sigma}$ & $f_{5\sigma}$ & $f_{5\sigma}$\\
         & [$\mu m$] & [$\mu Jy$]  & [$\mu Jy$] & [$\mu Jy$] & [$\mu Jy$] \\
         \hline
         PHI1\_1& 25.0 & 70.8 & 233.9 & 75.0 & 233.3 \\
         PHI1\_2& 27.8 & 78.7 & 249.2 & 83.1 & 258.5 \\
         PHI1\_3& 30.9 & 87.6 & 277.0 & 92.7 & 288.4 \\
         PHI1\_4& 34.3 & 99.0 & 312.9 & 102.9 & 320.1 \\
         PHI1\_5& 38.1 & 113.0 & 358.9 & 114.3 & 355.6 \\
         PHI1\_6& 42.6 & 134.2 & 424.2 & 126.9 & 394.8 \\
         PHI2\_1& 47.4 & 82.9 & 262.1 & 141.0 & 438.7 \\
         PHI2\_2& 52.3 & 94.3 & 298.2 & 156.9 & 488.1 \\
         PHI2\_3& 58.1 & 108.4 & 342.9 & 174.3 & 542.3 \\
         PHI2\_4& 64.5 & 123.2 & 389.6 & 193.5 & 602.0 \\
         PHI2\_5& 71.7 & 152.5 & 482.3 & 215.1 & 669.2 \\
         PHI2\_6& 79.7 & 172.1 & 544.3 & 239.1 & 743.9 \\
         PPI1   & 96.3 & 28.6 & 90.4 & 276.0 & 858.7 \\
         PPI2   & 126.0 & 44.9 & 142.0 & 378.0 & 1176.0 \\
         PPI3   & 172.0 & 67.0 & 211.7 & 516.0 & 1605.0 \\
         PPI4   & 235.0 & 81.6 & 258.2 & 705.0 & 2193.3 \\ 
    \end{tabular}
\tablefoot{In the table we report the PRIMAger filter bandpasses with their respective central wavelength and $5\sigma$ depths in the Deep and Wide surveys considered in this work. Both surveys correspond to 1500h, but spread over 1 $deg^{2}$ for the Deep and 10 $deg^{2}$ for the Wide. We report both the conservative expectation survey depths (conservative) and the guaranteed payload requirement depths (payload), both excluding confusion. Note that PHI provides $R=10$ continuous spectral coverage, which we here represent as discrete filters, or bands, to make the calculations more tractable.}
\end{table}

\begin{figure}
    \centering
    \includegraphics[width=\linewidth,keepaspectratio]{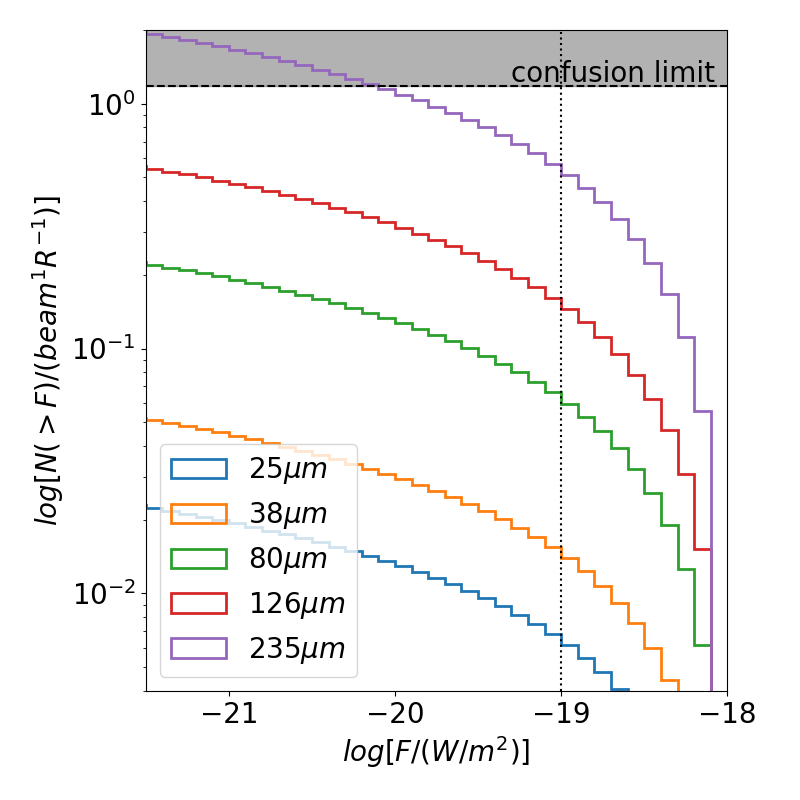}
    \caption{We do not expect FIRESS to be affected by line confusion down to at least $10^{-20}W/m^{2}$. We show the cumulative line number counts per beam and resolution element at different wavelengths, as derived using \spr. The vertical dotted line shows the flux of $10^{-19}W/m^{2}$, reachable with FIRESS with one hour of integration time. The horizontal dashed line and the shaded region show the limit of 15/beam/R, above which PRIMA could be affected by line confusion at the longest wavelengths. }
    \label{fig:Confusion_spectra}
\end{figure}

From these photometric surveys it will be possible to select a sub-sample of sources to follow-up spectroscopically with FIRESS. In particular, with one hour of integration time we expect to detected lines with fluxes above $10^{-19}\,W/m^{2}$ with the low-resolution mode of FIRESS. As shown in Fig. \ref{fig:Confusion_spectra}, we expect the number of observed spectroscopic sources to be well below the limit of 15 sources per beam per resolution element, above which we expect PRIMA could be affected by line confusion. We note that these line counts do not include the effects of confusion which could increase line blending in some cases. However, at the depth of $10^{-19}\,W/m^{2}$, FIRESS should not be largely affected by line confusion, even at the longest observed wavelength (i.e. 235$\,\mu m$). This is true even down to a depth of $10^{-20}\,W/m^{2}$.

\subsection{\spr}\label{sec:spritz}
The \spr simulation is particularly well suited to predictions for PRIMA as it starts from previously observed IR luminosity functions \citep{Gruppioni2013} of star-forming galaxies, AGN and composite systems. These are complemented by the K-band luminosity function of elliptical galaxies and the galaxy stellar mass function of dwarf irregulars. These functions are used in \spr to derive the number of galaxies expected at different redshifts ($z=0-10$) and IR luminosities ($log_{10}(L_{\rm IR} \rm {L_{\odot}^{-1}})=5-15$).
The galaxies included in the simulations can be broadly grouped into four populations:
\begin{itemize}
    \item Star-forming galaxies correspond to star-forming galaxies with no evident sign of AGN activity. This class includes spirals and starbursts, as derived from the observed \hers{} IR luminosity functions \citep{Gruppioni2013}, dwarf irregulars, derived from their observed galaxy stellar mass function \citep{Huertas-Company2016,Moffett2016}. Spirals have specific star-formation rates $\rm log({\rm sSFR}/{\rm yr^{-1}})=-10.4$ to $-8.9$, while starbursts have higher sSFR ranging from $\rm log({\rm sSFR}/{\rm yr^{-1}})=-8.8$ to $-8.1$. Dwarf irregulars are instead galaxies with lower stellar masses, as their characteristic stellar mass (i.e., mass at the knee of the mass function) is $log_{10}(M_{*}/M_{\odot})\leq11$.
    \item AGN-dominated systems are galaxies whose mid-IR emission is dominated by AGN activity. They include two different populations which differ by their optical extinction, i.e. AGN1 and AGN2. Their number densities have been derived by \citet{Bisigello2021} combining the observed AGN IR and UV observed luminosity functions \citep{Gruppioni2013,Croom2009,McGreer2013,Ross2013,Akiyama2018,Schindler2019}.  
    \item Composite systems are objects whose energetics are dominated by star formation, but which include a faint AGN component. They are split into star-forming AGN (SF-AGN), which have an intrinsically faint AGN (i.e., $L_{\rm BOL}\leq10^{13}\,{\rm L}_{\odot}$), and SB-AGN, whose AGN is bright but extremely obscured (i.e., $\log(N_{\rm H}/{\rm cm}^{-2})=23.5-24.5$). Both populations are derived starting from the observed \hers luminosity function \citep{Gruppioni2013}.
    \item Passive galaxies are elliptical galaxies, generally with little or no star-formation, as derived from the observed $K$-band luminosity functions \citep{Arnouts2007,Cirasuolo2007,Beare2019}. Some of these galaxies may host an obscured AGN.
\end{itemize}

To each simulated galaxy we assign one SED model that depends on its galaxy population \citep{Polletta2007,Rieke2009,Gruppioni2010,Bianchi2018} in order to derive the photometric fluxes expected in different filters, spanning from UV to radio wavelengths. The proportions of galaxies belonging to each population is derived by the luminosity function evolution of the different galaxy classes, at different redshifts and luminosities. We then fit the empirical SED assigned to each simulated galaxy using 
the software {\sc sed3fit} \citep[for AGN;][]{Berta2013}, in order to disentangle the AGN from the galaxy contribution, or the multi-wavelength analysis of galaxy physical properties {\sc Magphys} \citep[for non-active objects;][]{daCunha2008}. From the fit we retrieved the stellar mass, the accretion luminosity, the hydrogen column density of the dusty torus, and the contributions to the IR luminosity of star-formation and AGN activity. The SFR is directly taken from the UV continuum and the IR luminosity, for the unobscured and obscured components, respectively. We then used several theoretical and empirical relations to derive additional physical properties, which are marginally or not constrained from the SED fitting, such as the X-ray AGN luminosity and the metallicity.

The simulation also includes IR emission lines due to star-formation or AGN activity associated to the SEDs through several empirical relations \citep[e.g.][]{Bonato2019,Gruppioni2016}. In the next section we report some update required to take into account the dependence on metallicity of some key IR lines.

Overall, \spr{} is consistent with a large set of observations from $z=0$ to $z=6$. This includes luminosity functions and number counts at different redshifts from X-rays to radio, the total galaxy stellar mass function, the molecular gas mass density, the CO and [CII] luminosity functions \citep{Bisigello2022}, AGN diagnostic diagrams \citep{Bisigello2024}, and the SFR versus stellar mass plane. We refer to \citet{Bisigello2021,Bisigello2022} and \citet{Bisigello2024} for further comparisons with observations and more details on the \spr simulation. 

\subsubsection*{Linking far-IR lines to gas metallicity in \spr} \label{sec:spritz_lines}

\begin{figure*}
    \centering
    \includegraphics[width=\linewidth,keepaspectratio]{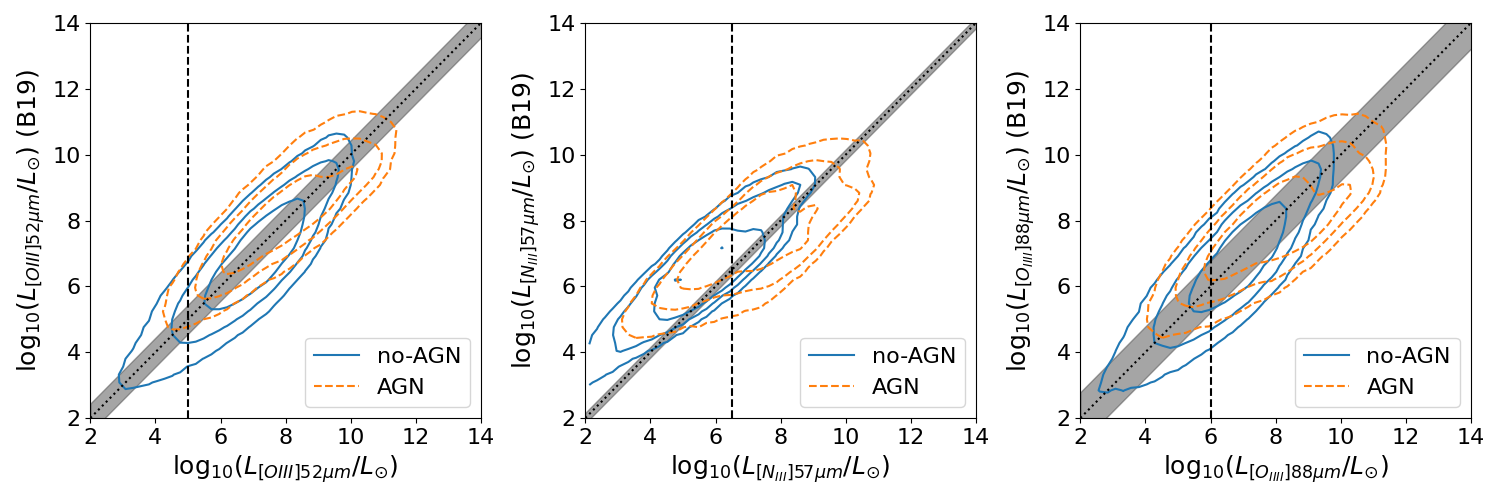}
    \caption{The metallicity-dependent method outlined in Sect. \ref{sec:spritz_lines} gives luminosity estimates consistent with those derived using the empirical relations by \citep{Bonato2019}. In the figure we show the comparison for the \oiiia (left), \niii (center) and \oiiib (right) lines. Dotted lines indicate a 1:1 relation, with the shaded region showing the 1$\sigma$ of the relations by \citep{Bonato2019}. The vertical dashed lines show the minimum line luminosity used in the relation by \citep{Bonato2019}. Contours correspond to 68\%, 95\% and 99.7\% of the distribution of objects hosting an AGN (orange dashed contours) and non active galaxies (solid blue contours).}
    \label{fig:line_comparison}
\end{figure*}

Rest frame mid- and far-IR emission lines are included in the \spr simulation using the empirical relations by \citet{Bonato2019} and \citet{Gruppioni2016}, linking the line luminosity to $L_{IR}$. However, future IR facilities, such as PRIMA, with sensitive spectral capabilities can take advantage of metal-sensitive far-IR lines including \niii, \oiiia and \oiiib to trace gas metallicity. For this reason we implement an alternative approach to estimate the strengths of these key far-IR lines in the galaxies modeled with \spr. \par

As demonstrated by \citet{Lamarche2022}, the absolute O/H abundance can be derived directly from the far-IR cooling lines \oiiia and/or \oiiib measurements, when supplemented with a tracer of hydrogen emission measure (recombination or free-free emission).  The \emph{relative} abundance measure N/O is also of strong interest, since nitrogen undergoes enhancement relative to oxygen from secondary production in CNO cycle stars. The N/O abundance has been established to correlate well with absolute abundance locally \citep{Nagao2011,Pereira-Santaella2017,Spinoglio2022}, revealing significant discrepancies with optical abundance measure in dusty local systems \citep{PengB2021,Chartab2022}.

To incorporate realistic luminosities for abundance-sensitive lines, we start by incorporating the \oiiib line using the analytical model by \citet{Yang2020}, which links the \oiiib luminosity directly to the gas metallicity (Z) and the SFR:
\begin{equation}
    \frac{L_{[\ion{\rm O}{III}]88\mu m}}{L_{\odot}}=10^{8.27-0.0029\,(7.3+log(Z/Z_{\odot}))^{2.5}}\,\frac{Z}{Z_{\odot}}\,\frac{SFR}{M_{\odot}/yr}
\end{equation}
where $Z_{\odot}$ indicated the solar value. 

The ratio between the \oiiia and \oiiib line is a strong function of the electron density $n_{e}$ \citep[e.g.,][]{Draine2011}. In \spr, we randomly assign an electron density $n_{e}$ between $10^{2}$ and $10^{3}\,cm^{-3}$ to each galaxy and derive the \oiiia expected luminosity directly from the \oiiib luminosity. These electron densities are consistent with measurements obtained using mid and far-IR density-sensitive tracers in local Universe systems spanning moderate to very high specific SFRs \citep[e.g.,][]{Dale2006,Inami2013,HerreraCamus2016}. 

From the metallicity and the respective oxygen abundance, we retrieved the expected nitrogen abundance, as quantified in the local universe using observations of late type galaxies  by \citet{Pilyugin2014} :

\begin{align}
    12 + log(N/H)= 2.47 (12 + log(O/H))-&13.43 \\ 
    &\text{if } 12 + log(O/H)\geq8.1 \nonumber \\ 
    12 + log(N/H)= 0.96 (12 + log(O/H))-&1.20 \nonumber \\
    &\text{if } 12 + log(O/H)<8.1 \nonumber 
\end{align}
We then made use of this information to estimate the nitrogen to oxygen abundance ratio (N/O). This allowed us to predict the \niii luminosity from the \oiiia line, assuming an electron density of $10^{4} cm^{-3}$\ %
(near both lines' critical densities) %
and a temperature of $10^{4}\,K$ 
\begin{equation}
    \frac{L_{[\ion{\rm N}{III}]57\mu m}}{L_{[\ion{\rm O}{III}]52\mu m}}=\frac{N}{O}\,\frac{1}{0.4}\,\frac{1.0+0.377\,x+0.0205\,x^{2}}{1.0+0.691\,x+0.0966\,x^{2}}
\end{equation}
where $x=n_{e}/\sqrt{T}$ \citep{PengB2021}.

The above methods allow us to derive the line luminosity powered by stellar activity, however, all three lines can be also powered by AGN activity. In order to include this contribution, we used the models from \citet{Feltre2016}, computed with the photoionization code {\sc CLOUDY} \citep[v13.03,][]{Ferland2013}. In particular, we consider a ionisation parameter at the Str\"omgren radius randomly varying between log$_{10}( U_{\rm S})$=-1.5 and -3.5, a range of sub- and super-solar interstellar metallicity (0.008, 0.017, 0.03), a UV spectral index $\alpha=$-1.4, a dust-to-metal ratio of 0.3 and internal micro-turbulence velocity v$=$100km/s \citep[see][]{Mignoli2019}. The hydrogen number density is assumed to be 10$^{3}$ cm$^{-3}$.

In Figure \ref{fig:line_comparison} we show the comparison between these line flux estimates and the empirical relation presented in \citet{Bonato2019}. The estimates for \oiiia and \oiiib are largely consistent with the empirical relations.  The estimated \niii line luminosity, however, while consistent for AGN and composite objects, is on average around 1 dex below the empirical relation for purely star-forming objects. These line luminosities, however, are well below the luminosity range traced by the observations used in \citep{Bonato2019}. Similar results, although with a larger scatter, are obtained if we consider the line luminosities obtained using the empirical relations by \citep{Gruppioni2016}. We conclude therefore, that for the purposes of predicting PRIMA detection rates, the estimated line luminosities of \oiiia, \oiiib and \niii in \spr used here, which include a realistic dependence on metallicity, are generally consistent with previous observations.

\subsection{Santa Cruz SAM}\label{sec:sam}

In addition to \spr, we have made use also of the state-of-the-art Santa Cruz semi-analytic model \citep[SC SAM][]{{Somerville2015}}. For the purpose of this work, we utilised five simulated past lightcones presented by \citet{Yung2023}, each covering an area of 2 deg$^2$ and spanning a redshift range of $0 < z \lesssim 10$. These lightcones are constructed with halos extracted from the dark matter-only cosmological SMDPL simulation from the MultiDark suite, which has a volume of (400 Mpc $h^{-1}$)$^3$ and reliably resolves halos down to $M_\text{h} \sim 10^{10}$ M$_\odot$ \citep{Klypin2016}. For halos in a lightcone, Monte Carlo realisations of dark matter halo merger histories are constructed using an extended Press-Schechter (EPS)-based algorithm \citep[e.g.][]{Somerville1999a}, which provides input for the SC SAM described below. We refer the reader to \citet{Yung2022} and \citet{Somerville2021} for details regarding the constructions of lightcones.

The galaxies and AGN in the lightcones are simulated with the versatile and well-established SC SAM \citep{Somerville1999, Somerville2008, Somerville2015}, which models the formation and evolution of large ensembles of galaxies by tracking the flows of mass and metals into and out of the intergalactic medium, galaxy halos, and galaxies. These flows are driven by physical processes occurring across a vast range of scales, including gas accretion from the cosmic web into halos, cooling from the circumgalactic medium into cold gas reservoirs within galaxies, BH accretion in the centres of galaxies, and large-scale galactic outflows driven by supernovae and AGN feedback. The SAM provides predictions for a wide range of physical properties, including stellar mass, SFR, black hole mass, and BHAR \citep{Somerville2021,Yung2019}. 
The simulated galaxy populations in the lightcones are complete down to $M_*\sim 10^7$ M$_\odot$.
With physical parameters calibrated to a subset of observed galaxy properties at $z\sim0$, the SAM is able to reproduce the evolution of many observational quantities, including multi-wavelength luminosity functions up to $z\sim10$, one-point distribution functions of stellar mass and SFR up to $z\sim8$, and various scaling relations \citep{Somerville2015, Somerville2021, Yung2019, Yung2019b}. The spatial distribution  of sources in the lightcone are in excellent agreement with the observed two-point angular correlation functions \citep{Yung2022}. In addition, the model has also been shown to reproduce the observed evolution in AGN bolometric, UV, and X-ray luminosity functions up to $z \sim 5$ \citep{Hirschmann2012, Yung2021}.


\subsection{Deriving SFR and BHAR from far-IR spectro-photometric data}\label{sec:sed-dec}
\begin{figure}
    \centering
    \includegraphics[width=\linewidth,height=7cm]{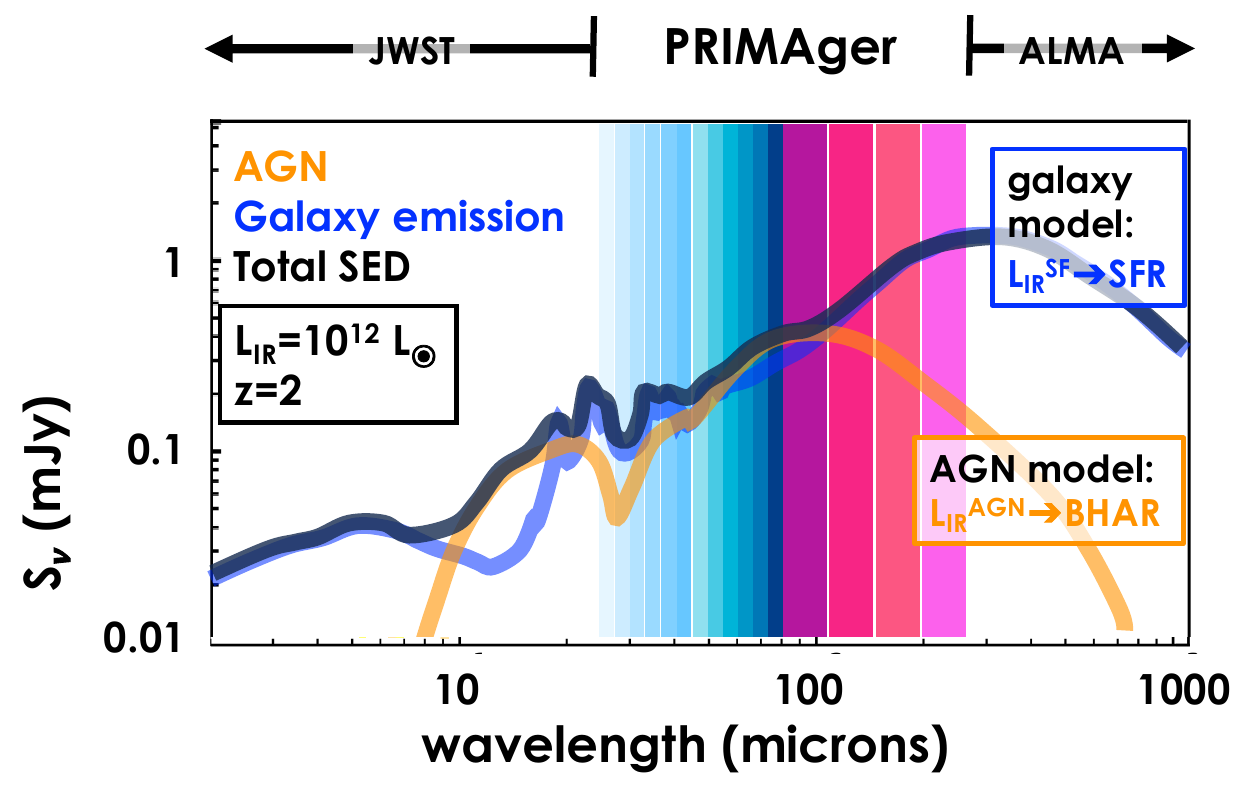}
    \caption{Example of SED decomposition for an L$_{\rm IR}$=10$^{12}$ galaxy at $z$$=$2. Star formation (blue), and AGN heated dust (orange), are shown schematically here and combine to produce a composite galaxy SED (black) that can be decomposed using the PRIMAger bands, shown here in different colours, from pale blue to magenta.} 
    \label{fig:SED_decompose}
\end{figure}

 \begin{figure*}
\sidecaption
    \includegraphics[width=12cm,keepaspectratio]{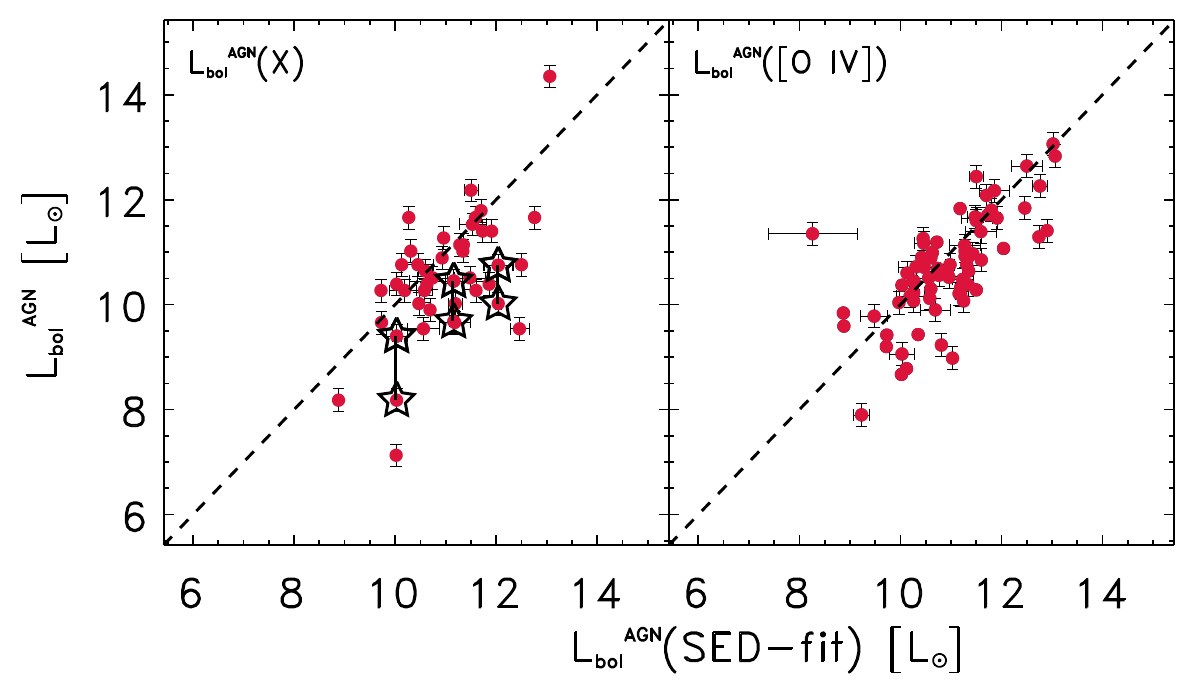}
    \caption{The power of SED decomposition to derive the AGN bolometric luminosity ($L_{bol}^{AGN}$), and the BHAR. The panels shows the AGN bolometric luminosity derived from SED fitting by \citet{Gruppioni2016} for a local sample of AGN versus the AGN bolometric luminosity derived from different indicators: {\em left}: X-ray (2-10 keV) Luminosity; {\em right}: [O~IV] $25.9\mu m$ far-IR line. X-ray bolometric correction are taken from \citet{Brightman2011}. The dashed line represents the 1-1 relation. With PRIMA we will be able to perform similar decompositions at least up to $z=2$. The stars in the left panel show how the X-ray 
    derived $L_{\rm bol}^{AGN}$ increases for Compton-thick AGN once the hard X-ray spectrum is included \citep[as shown by][]{LaCaria2019}. The correlation found in the right panel can be used as a calibration for the SED-fitting results when far-IR spectroscopy is not available.}
    \label{fig:LbolIR_LbolX}
\end{figure*}

SEDs that extend from approximately $15$~$\mu$m to $150$~$\mu$m in the rest frame can be used to reliably determine the respective importance of star formation and black hole accretion activity \citep{Kirkpatrick2015,Ciesla2015,Gruppioni2015}, as shown in Fig. \ref{fig:SED_decompose}. This tends to work because the AGN contributes primarily to warm dust that emits in the rest frame mid-IR range peaking around 20~$\mu$m, while star formation activity creates a colder dust component which dominates the far-IR, peaking around $100$~$\mu$m \citep{Shimizu2017}. 

Studies based on SED decomposition show that the warm dust temperature, the colour temperature and the relative luminosity of the cold and warm components are all excellent indicators of the fraction of power contributed by the AGN in the mid-IR and bolometrically \citep{Kirkpatrick2015}. For highly obscured galaxies, such as those prevalent at $z\sim1-2$, dust luminosity is an excellent proxy for the bolometric luminosity of the system. Thus warm dust mid-IR luminosities can be used to directly compute the bolometric luminosity of the AGN ($L_{\rm bol}^{\rm AGN}$). Indeed, as shown for example by \citep{Gruppioni2016} and reported in Fig. \ref{fig:LbolIR_LbolX} using a sample of local AGN, the AGN bolometric luminosity derived through SED-fitting and decomposition is in very good agreement with the AGN bolometric luminosity derived from high ionization fine structure lines such as \oiv. We note that for the data shown in the Figure a linear regression fit gives a slope of $0.9\pm0.3$ for the [O IV] vs. SED-fit derivations, with the best-fit line crossing the 1-1 relation at $L_{bol}(\rm SED-fit)\sim10^{10}\,L_{\odot}$. However, a more accurate analysis should be done on larger and complete samples in order to be used as calibration. 
On the contrary, X-rays tend to underestimate the AGN luminosity, particularly for highly obscured objects. Indeed, the X-ray bolometric corrections of Compton-thick AGN (empty stars in Fig. \ref{fig:LbolIR_LbolX}) derived by \citet{Brightman2011} using 2-10 keV X-ray observations increased once hard X-ray measurements became available \citep[e.g., NuSTAR][]{LaCaria2019}, moving it closer to the estimation from SED fitting. The possibility that the absorption-corrected 2-10 keV luminosity, and consequently $L_{BOL}(X)$, is underestimated for moderately obscured/Compton-thick AGN, if additional multi-wavelength data are not taken into account (e.g., SED-fitting), has also been found by \citet{Lanzuisi2015}.

Overall, these correlations show the capability of the SED decomposition to derive $L_{\rm bol}^{\rm AGN}$ for sources without mid-IR spectroscopy (i.e., too faint or high-$z$). Once the AGN bolometric luminosity is known, it is possible to compute the black-hole accretion rate, BHAR, \citep[through the standard assumption of 10\% efficiency,][]{Hopkins2007}. Similarly, SFRs can be derived from the colder dust component visible in the far-IR (blue curve in Fig. \ref{fig:SED_decompose}) using the standard relations derived for local galaxies \citep{Kennicutt2012} and bright lines such as \neii \citep[e.g.][]{Spinoglio2022}.

\section{Results}\label{sec:results}

The blending of sources due to the limited spatial resolution of far-IR telescopes can be an important limitation to the ultimate attainable sensitivity in an observation. This limit is usually characterized by a so-called ``confusion noise,'' which corresponds to the level of the blended background of sources at a given wavelength and resolution.   Recent work demonstrates that deblending techniques based on the Bayesian source extractor {\tt XID+} can be used to push below the classical confusion (the confusion limit reached by a basic blind source extractor) in PRIMAger \citep{Donnellan2024}, by leveraging the information provided by shorter wavelengths and the dense wavelength coverage of the instrument. In particular, \citet{Donnellan2024} show that a self-contained approach based on using positional priors from a source catalog detected in the shortest PRIMAger bands with Wiener-filtering to deblend the longer PRIMAger wavelengths can reach factors of $2-3$ below the classical confusion limit. The use of a denser catalog (for example, obtained from \textit{Roman} Space Telescope IR observations of the same field) together with weak intensity priors can push this limit even further, factors of $5-10$ at the longer wavelengths. Most importantly, \citet{Donnellan2024} show that the self-contained Wiener-filter catalog approach allows PRIMAger to recover the SEDs of galaxies at the knee of the luminosity function at $z=2$ out to PPI\_2 (see their figures 7 and 9). The calculations presented in the following sections use these results.

\subsection{The galaxy population observed by PRIMAger} 
\begin{figure}
    \centering
    \includegraphics[trim=20 5 20 20,width=\linewidth,keepaspectratio]{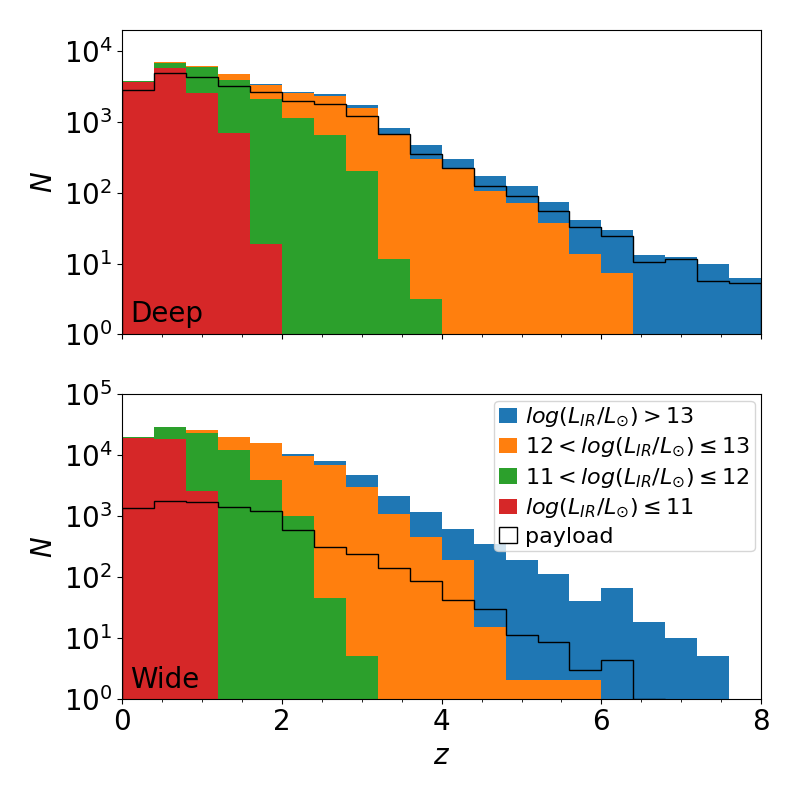}
    \caption{The proposed PRIMAger surveys will not be limited to only the brigthest IR galaxies. We show the number of galaxies as a function of redshift for the Deep (top, 1500$\,h/deg^{2}$) and Wide (bottom, 150$\,h/deg^{2}$) surveys using the \spr simulation. We considered only galaxies with $S/N\geq5$ in at least 6 out of 12 PRIMAger Hyperspectra filters (i.e., PHI1-PHI2 in Table \ref{tab:depths}). The black solid lines show the total redshift distributions considering the payload depths.}
    \label{fig:zdistribution}
\end{figure}

As shown in Fig. \ref{fig:zdistribution}, using the conservative depths we can detect galaxies with $L_{IR}>10^{13}\,L_{\odot}$ up to $z=7-8$ in both the deep and wide PRIMAger surveys. Less-extreme galaxies with luminosities down to $L_{IR}=10^{11}\,L_{\odot}$ can be detected to $z=4$ in the Deep survey and up to $z=3$ in the Wide survey. As a comparison, with \hers it was possible to observe galaxies with $L_{IR}=10^{11}\,L_{\odot}$ only to $z<1$ \citep[e.g.,][]{Gruppioni2013,Magnelli2013}. If we consider the payload depths, which are the minimal guarantee depths, we loose $28\%$ galaxies in the Deep and $35\%$ in the Wide survey, but in the latter galaxies are limited to $z=6.5$.
\par

\begin{figure}
    \centering
    \includegraphics[trim=40 40 40 0,width=0.9\columnwidth,keepaspectratio]{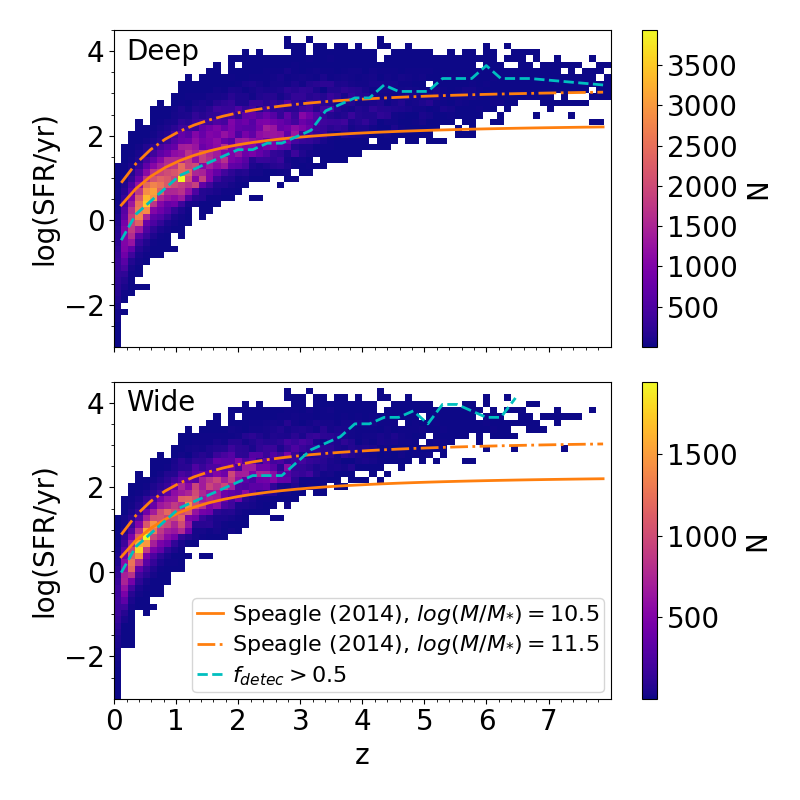}
    \caption{PRIMA surveys will not be limited to extremely bright objects, but will start to probe normal star-forming galaxies. We show the distribution of SFR as a function of redshift for the Deep (top) and Wide (bottom) surveys, using the conservative depths (see Table~\ref{tab:depths}) and the \spr simulation. The cyan dashed lines indicate the 50\% completeness level, while the orange solid and the dash-dotted lines show the SFR of a typical main sequence star-forming galaxy with $\rm log(M/M_{*})=10.5$ and $\rm log(M/M_{*})=11.5$, respectively. These have been derived considering the $\rm M_{*}-SFR$ relation by \citet{Speagle2014} at different redshifts.}
    \label{fig:SFR_z}
\end{figure}

To understand the physical properties of these photometric samples, we present in Fig. \ref{fig:SFR_z} the SFR as a function of redshift for both the Deep and the Wide PRIMAger surveys. Using the conservative depths, we expect the Deep survey to be $50\%$ complete for star-forming galaxies with stellar masses $\rm log(M_{*}/M_{\odot})=10.5$ and 11.5 up to $z=2.3$ and $z=4$, respectively. Note that for galaxies over $\rm log(M_{*}/M_{\odot})=10.5$ at $z\sim2$ the study by \citet{Donnellan2024} shows that the SEDs are well measured over a large fraction of the bandpass using the {\tt XID+} deblending (see their Figure 9), so we expect to be able to accurately measure their SFR and BHAR.
The Wide survey should be 50\% complete for galaxies with $\rm log(M_{*}/M_{\odot})=11.5$ up to $z=2.9$. Photometric surveys with PRIMAger would not be limited only to the brightest, most luminous IR galaxies at $z > 1$, as was the case for previous far-IR missions, but would finally start to also include a more representative sample of star-forming galaxies out to cosmic noon and beyond.

\subsection{SFR vs. BHAR with PRIMAger using the far-IR continuum} 
As previously discussed, the rest-frame continuum between 15 and 150 $\mu m$ can be used to disentangle the star formation and black hole accretion activity, as the dust heated by the these mechanisms has two different temperature distributions. In the next sections we discuss the accuracy of these AGN-SF decomposition using PRIMAGER bands and how we can use these estimates to improve galaxy evolution models.

\subsubsection{Parameter Extraction from SEDs}
\label{sec:extraction}

\begin{figure}
    \centering
    \includegraphics[width=\linewidth,keepaspectratio]{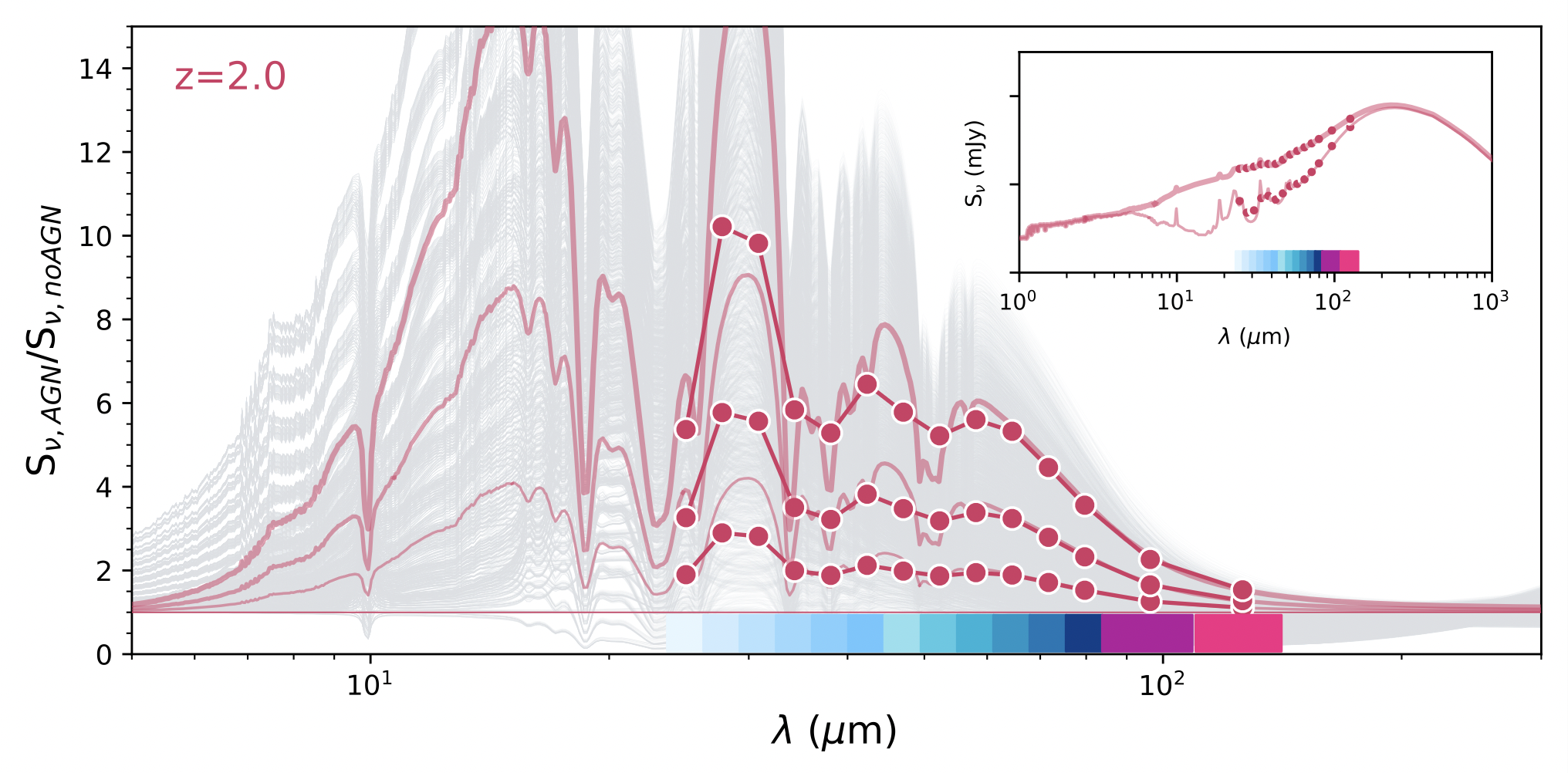}
    \caption{Contribution of AGN emission into the PRIMAger bandpass, showing that the PRIMAger bands are very sensitive to the presence of an AGN. Thin grey lines show a $z=2$ galaxy SED including different AGN properties (see Table \ref{tab:mocks}), divided by the SED without an AGN component. As an example, the emission of one AGN model with different $f_{\rm AGN}$ contributions (15, 30, and 45\% of the total IR luminosity) is shown in red thick lines. The fluxes obtained for integration in the equivalent filters with center wavelengths listed in Table \ref{tab:depths} for PHI+PPI1+PPI2 are shown as red circles.  The inset panel shows the SEDs corresponding to the red lines (for $f_{AGN}=0\%$ and $f_{AGN}=45\%$) over the full UV to submm wavelength range. The PRIMAger bands are well suited to probe the impact of AGN emission on the IR SED of galaxies.}
    \label{fig:SED_primagerfilters}
\end{figure}

While the mid- and far-IR continuum of galaxies can be used to obtain a simultaneous estimate of both the SFR and the BHAR, the quality of such estimations are correlated with the number of filters used for the SED decomposition and their wavelengths. The PRIMAger PHI continuous wavelength coverage $\lambda=25-80$\,$\mu$m and the PPI filters at $\lambda=96$ to 235 $\mu$m efficiently probe the presence of PAHs for $z\gtrsim1$ and separate the AGN contribution, peaking at $\lambda_{rest}\sim20-30$~$\mu$m, from the emission powered by star-formation, which peaks at $\lambda_{rest}\sim60-120$~$\mu$m (Fig. \ref{fig:SED_decompose}).
In this section, we test the ability to recover physical parameters from the SED information provided by PRIMAger. Specifically, we simultaneously extract the fraction of the total grain mass contributed by PAHs (q$_{PAH}$), the total IR luminosity (L$_{IR}$), and the fraction of the total IR luminosity due to an AGN (f$_{AGN}$) using SED modeling.

We use \texttt{CIGALE}\footnote{\url{https://cigale.lam.fr/}} \citep{Boquien2019} to compute multi-wavelength SEDs of AGN host galaxies \citep{Ciesla2015,YangG2020,YangG2022}. This analysis is performed in a self-consistent way (i.e. creating and analysis the models with the same tool) and over a grid of SED templates wider that the one included in \spr, to avoid introducing any systematic uncertainties due to the SED fitting procedure or the simulation. In particular, the \texttt{skirtor} library \citep{Stalevski2016} is used to model the AGN emission. Dust emission is modelled with the library of \cite{Draine2014}, an extended version of the models of \cite{DraineLi2007}. We create a set of AGN+host galaxy SEDs at different redshifts (our mock catalogue), varying dust emission and AGN properties (see Table~\ref{tab:mocks}), including the contribution of the AGN to the total L$_{IR}$ (f$_{AGN}$) to create a wide range of SED types. 
The SEDs are built with $q_{PAH}$ values varying from 0.47\% to 7.32\% to probe a wide range of SEDs.
\cite{Schreiber2018} showed that $q_{PAH}$ can vary from approximately 7\% down to 2.5\% between $z=0$ and $z=3$. 
Furthermore, in the local universe, typical normal star-forming galaxies have a $q_{PAH}$ ranging from 1.22 to 4.58\% \citep{Ciesla2014}.
Therefore, the dynamical range of $q_{PAH}$ probed in this test covers typical values found in the literature.
Regarding the AGN contribution, we use $f_{AGN}$ ranging from 0. to 0.5 which is consistent with the range of AGN fractions measured by \cite{Malek2018} when fitting \textit{Herschel} IR galaxies. For reference, the range of f$_{AGN}$ explored here corresponds to a range of f$_{AGN,MIR}$ from 0--90\% \citep{Kirkpatrick2015}. 

In Figure~\ref{fig:SED_primagerfilters} we show the variety of these mock SEDs and how the AGN contribution is well probed by the wavelengths covered by PRIMAger, using as example an object at $z=2$. The faint gray lines show the SED of our mock galaxies divided by that of a galaxy of the same luminosity without an AGN component, highlighting the fact that the effect of an AGN is significant at the wavelengths probed by the PHI. 

\begin{table}
    \caption{Parameters of the models used to build the mock SEDs with \texttt{CIGALE}.}
    \begin{center}
    \begin{tabular}{lll}
    \hline
    \hline
    Parameter     & Value           &                        \\ 
    \hline
    \hline
    \multicolumn{3}{c}{Dust emission: \citep{Draine2014}}\\ 
    \hline
    $q_{PAH}$ & 0.47, 1.12, 1.77, & Mass fraction of PAH   \\
              & 2.50, 3.19, 3.90,  &    \\
              & 4.58, 5.26, 5.95, &    \\
              & 6.63, 7.32  &    \\
    $\alpha$ & 2.5  &  Slope $dU/dM \propto U^{\alpha}$\\
    $\gamma$ & 0.001, 0.02, 0.2 & Fraction illuminated from   \\
             &                  & min. to max. radiation field  \\
    $U_{min}$ & 2, 10, 20, 30 & Minimum radiation field \\ 
    \hline
    \hline
    \multicolumn{3}{c}{AGN emission: \texttt{SKIRTOR} \citep{Stalevski2016} }                        \\ 
    \hline 
    $t$ & 3, 5, 7 & Opt. depth at 9.7$\mu$m\\
    $i$ & 30, 60 & Viewing angle  \\
    $f_{AGN}$& [0;0.5] & Contribution of the AGN  \\
             &  by steps of 0.05 & to the total L$_{IR}$  \\ 
    \hline
    \hline
    $z$    & 1, 2 & Redshift\\
    \hline
    \end{tabular}
    \end{center}
\label{tab:mocks}
\end{table}

We associate flux errors to the modelled fluxes, considering both the contribution from the instrumental (Table ~\ref{tab:depths}) and the confusion noise.
We use the PHI1\_4 filter ($\lambda=34$\,$\mu$m) as the detection band, because observations at that wavelength will not suffer from confusion \citep{Bethermin2024}, and compute the fluxes in the remaining filters from their ratio to PHI1\_4, while maintaining the overall SED shape.
Classical confusion noise is usually derived for a blind detection strategy at a single wavelength, but PRIMAger's broad and dense wavelength coverage enables more refined strategies that overcome such limitations. \citet{Donnellan2024} show that a self-contained Bayesian deblending approach implemented using \texttt{XID+} \citep{Hurley2017} reaches flux densities a factor of $\sim2-3$ fainter than the classical confusion limit estimated in \citet{Bethermin2024}. We thus take the confusion noise values found by \citet{Donnellan2024} using their Wiener-filter derived catalogue, add them in quadrature to the instrumental noise to determine the total noise, and require a S/R of 10 in the PHI1\_4 filter for our recovery experiment. We also compute the results requiring S/R=5 in PHI1\_4, which show a mild degradation and are in general intermediate between the extraction with PPI1+PPI2 and just PPI1 alone. We also show the result of a run using S/R=10 and the lower confusion noise obtained in XID+ extractions when using a deep catalog with some prior intensity information \citep{Donnellan2024}.  

\begin{figure*}
    \centering
    \includegraphics[trim=20 5 20 20,width=\textwidth,keepaspectratio]{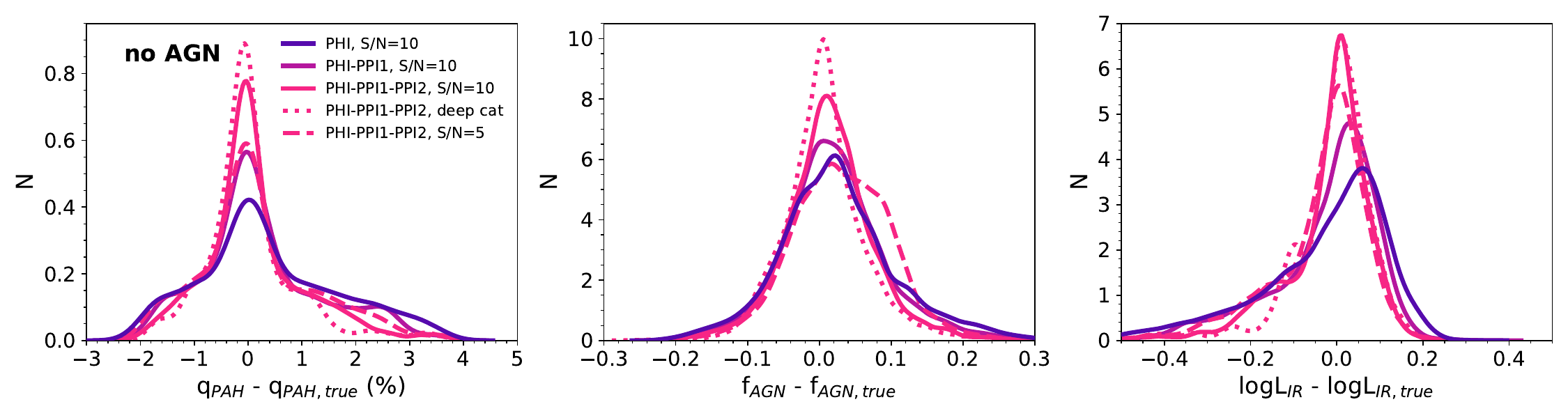}
    \caption{Normalised distributions of the difference between the estimated and the true values of the extracted parameters. In each panel, the lines are colour-coded according to the set of filters used in the fitting. The distributions include all models whose parameters are described in Table~\ref{tab:mocks}. The same distributions, with the PHI+PPI1+PPI2 set of filters but assuming a S/N of 5 instead of 10 in PHI1\_4 are shown in dashed line. The dotted line shows the results using the confusion noise for the deep catalog from \citet{Donnellan2024}, attainable in regions where ancillary information exists (e.g., from the \textit{Roman} Space Telescope). Left panel: PAH fraction measured in percent ($q_{PAH}$). Centre panel: AGN fractional contribution to the total IR luminosity ($f_{AGN}$). Right panel: total IR luminosity ($\log L_{IR}$). The distributions do not show any bias for the estimate of the PAH fraction and the AGN fraction at $z=1$ and $z=2$. The $\log L_{IR}$ estimated with only the PHI presents a bias of $+0.07$\,dex and a tail toward larger underestimates (factors of 2 to 3). Dropping the required S/R from 10 to 5 in the PHI1\_4 has only a very modest impact in $q_{PAH}$ and $f_{AGN}$ and does not introduce any biases.
}
    \label{fig:fagn_qpah_lir}
\end{figure*}

To test the quality of the parameter extraction we fit the mock catalogue for $q_{PAH}$, $f_{AGN}$, and $L_{IR}$ simultaneously, using the Bayesian-like analysis performed in \texttt{CIGALE}. We first do a run including data from PHI, then a run using PHI+PPI1, and a final run using PHI+PPI1+PPI2.
In Fig.~\ref{fig:fagn_qpah_lir}, we show the distribution of the difference between the physical properties estimated through this process and their true value, considering galaxies with redshifts of 1 and 2. Broadly speaking the histograms show very little or no bias for $q_{PAH}$ and $f_{AGN}$, while there is a mild bias in $L_{IR}$ when the PPI filters are not included. As expected, the histograms also tighten up when more wavelength information is included, and they show little to no degradation if the S/R is dropped from 10 to 5 at our anchoring filter PHI\_4.

Quantitatively, the histograms show a number of trends. Because the PAH emission is weakened or erased from the SED in the presence of a strong AGN continuum, we test the recovering of $q_{PAH}$ in cases where there is no AGN contribution.
The impact of reducing the wavelength coverage is on the width of the distribution, going from a dispersion of 0.9 with the full PHI plus two first filters of PPI to a flatter distribution when only using PHI (dispersion of $1.3$).
The AGN fractional contribution to the total $L_{IR}$ ($f_{AGN}$) is well recovered when the full PHI band and the two first PPI filters are used: the distribution is centred on zero and has a dispersion of 0.06. The bias introduced by removing the PPI filters is very slight, just 0.02, and the dispersion increases by 30\% to 0.08.
Finally, the distribution of the difference between the estimated $L_{IR}$ and its true value, in logarithmic scale, is centred on zero and has a dispersion 0.1\,dex when all filters are included, but the median is biased high to $+0.05$ with a dispersion of 0.15\,dex when we rely on PHI only.

\begin{figure}
    \centering
    \includegraphics[width=\columnwidth]{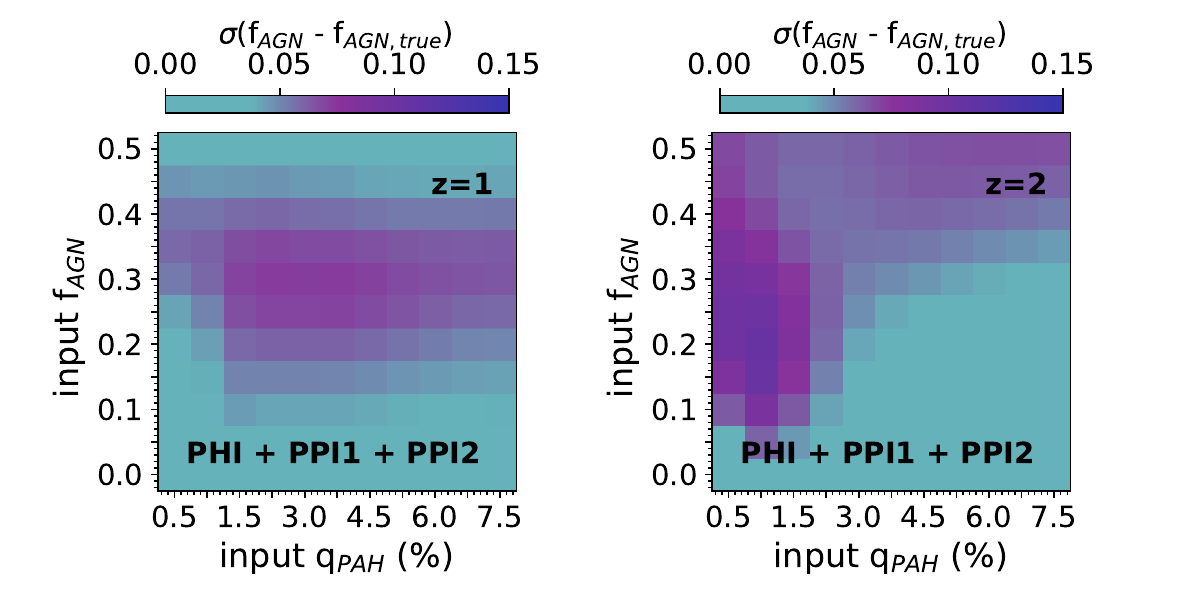}
    \includegraphics[width=\columnwidth]{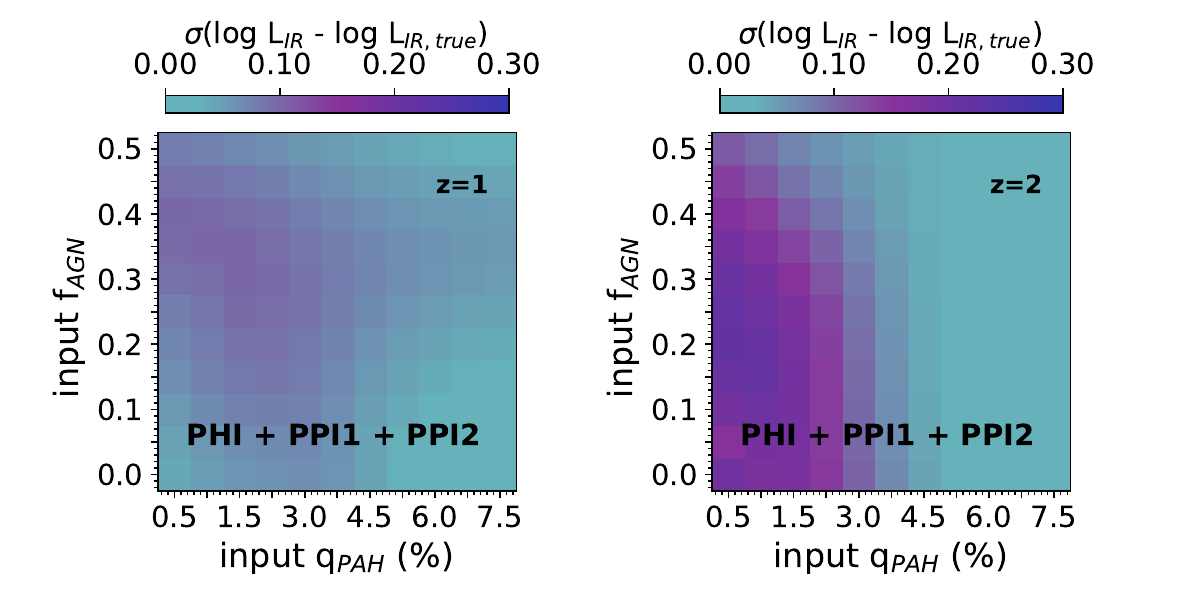}
    \caption{Extraction of $f_{AGN}$ and $L_{IR}$ as a function of AGN contribution and PAH fraction. The AGN fraction is well recovered over all the parameter range probed by the test, with an error $<0.1$ at both $z=1$ and $z=2$. The IR luminosity is recovered to better than 0.12\,dex, at $z=1$ and with a dispersion lower than $<0.25\,$dex at $z=2$. Top row panels: panels are colour-coded according to the standard deviation of the $f_{AGN}-f_{AGN,true}$ distribution. Left panel and right panels present the results at $z=1$ and $z=2$, respectively. Bottom row panels: same as top row panels for the $L_{IR}$. }
    \label{fig:fAGN_lir}
\end{figure}

It is also interesting to look at how the extraction results for $f_{AGN}$ and $L_{IR}$ vary as a function of both $q_{PAH}$ and $f_{AGN}$. We show the systematic error and dispersion in our extraction in Fig~\ref{fig:fAGN_lir} computed for $z=1$ and $z=2$, when employing a filter set that includes the full PHI, and PPI1 + PPI2. The dispersion of the $f_{AGN}-f_{AGN,true}$ distribution is on average 0.05, and below 0.12 over most of the parameter space. The highest uncertainty is found at $z=2$ for very low intrinsic PAH abundance, $q_{PAH}<2\%$, and intermediate AGN contributions, $0.15<f_{AGN}<0.35$.
These tests show that the PRIMAger bands are well-suited to probe the AGN emission and disentangle it from the emission due to star-formation for a vast range of source properties.
The bottom row of Fig.~\ref{fig:fAGN_lir} shows that the $L_{IR}$ is very well recovered at $z=1$ with a dispersion on the $\log L_{IR}-\log L_{IR,true}$ distribution well below 0.12\,dex for all the parameter space explored. 
At $z=2$ the filters probe shorter rest-frame wavelengths, further away from the peak in the far-IR, but the combined use of the PHI and the two first filters of PPI still allows for good recovery of the $L_{IR}$ with an average dispersion of 0.1\,dex and a maximum of 0.25\,dex.

\begin{figure}
    \centering
    \includegraphics[width=\columnwidth]{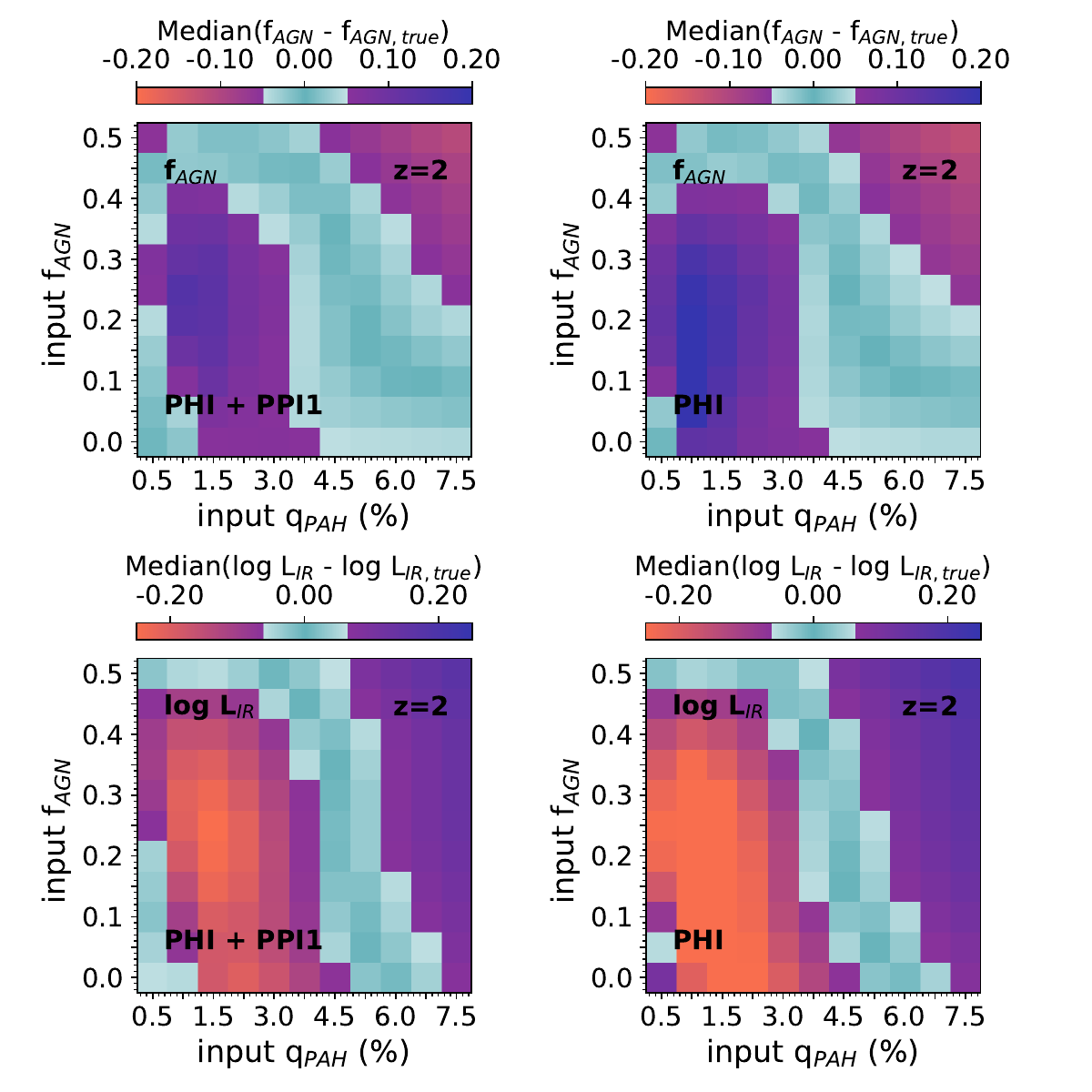}
    \caption{Impact on parameter extraction accuracy of removing long wavelength information at $z=2$. The inclusion of the longer wavelengths improves the accuracy of the determinations, but the degradation is fairly smooth. Top row: median value of $f_{AGN}-f_{AGN,true}$ as a function of $f_{AGN}$ and $L_{IR}$. Bottom row: same but for $\log L_{IR}-\log L_{IR,true}$. The left side and right side panels use the PHI band plus PPI1 and only in the full PHI band, respectively, to extract the parameters.}
    \label{fig:less_filters}
\end{figure}

To understand the impact of the longest wavelength SED points on the accuracy of the extracted parameters, we show in Fig.~\ref{fig:less_filters} the median value of the distributions shown in Fig.~\ref{fig:fagn_qpah_lir} and Fig.~\ref{fig:fAGN_lir} for $f_{AGN}$ and $L_{IR}$ at $z=2$, after removing one or both PPI filters. When removing the PPI2 filter (left column), $f_{AGN}$ is still well recovered with a bias that is lower than 0.05 in half of the parameter space tested here, except in the region of weak PAH and intermediate AGN contribution where it can reach up to 0.15, while for the $L_{IR}$ the bias can reach up to 0.25\,dex. If only the full PHI band is used (right column), biases for both $f_{AGN}$ and $L_{IR}$ also reach up to 0.15 and 0.25\,dex, respectively, but on a wider proportion of the parameter space. Therefore we conclude that the inclusion of the PPI fluxes are necessary to derive accurate PAH, SFR and AGN parameters, particularly for sources at $z=2$ where the PHI band probes only up to 27\,$\mu$m rest-frame wavelengths.

\subsubsection{Expected Distributions}

\begin{figure*}
\includegraphics[width=\textwidth]{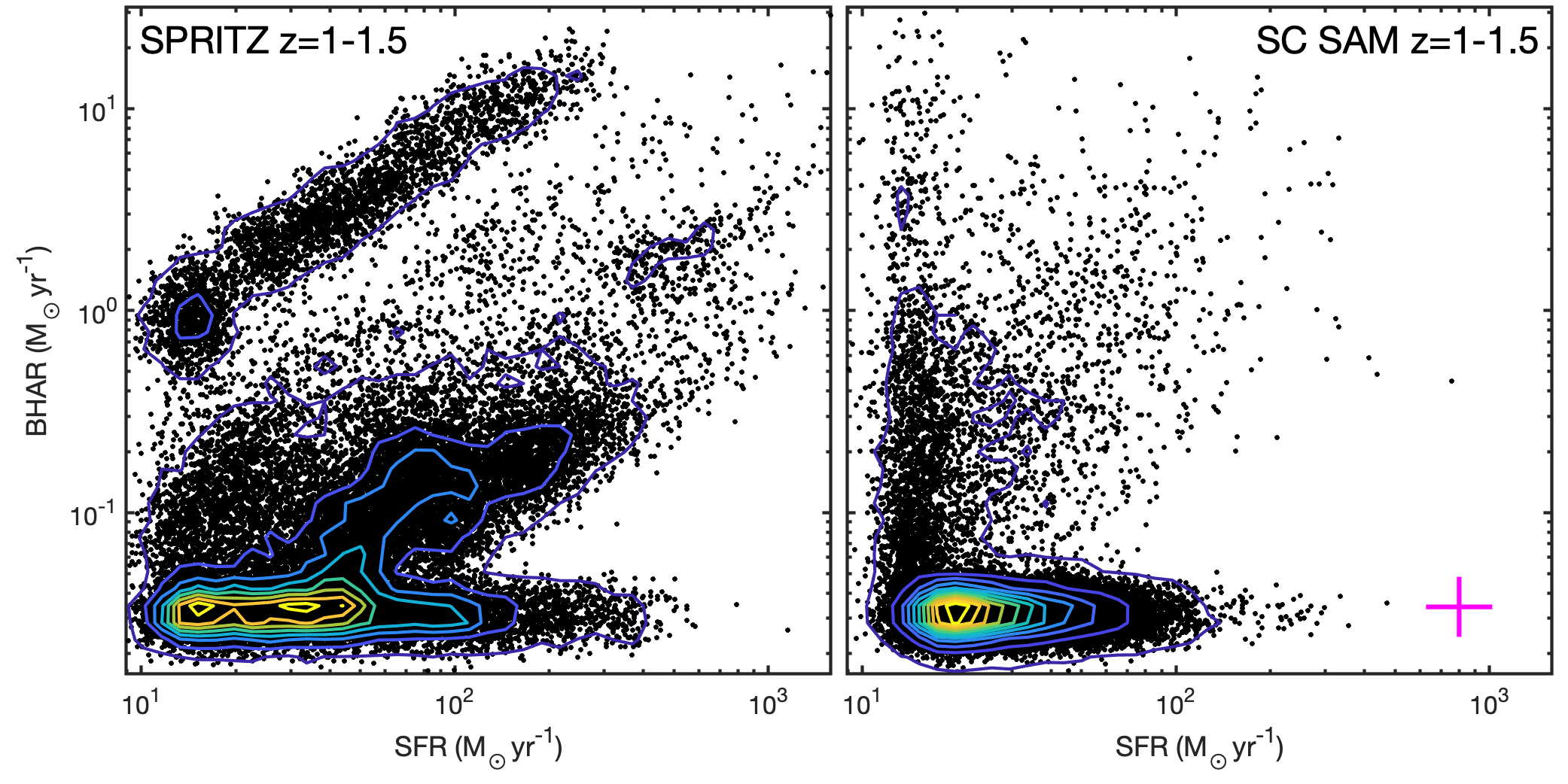}
\caption{Comparison of expected galaxy distributions in SFR and BHAR from \spr and the Santa Cruz SAM over a 10 sq. deg. light-cone with an integration time of $\sim150$~hrs. per square degree. The contours indicate density of sources, and the magenta cross in the right panel shows the expected extraction uncertainties for the median galaxy (\S\ref{sec:sed-dec}). Each panel has about 30,000 sources detected at 5$\sigma$ or better over half the PHI bandpass, but only $\sim4\%$ are hybrid AGN+SF galaxies in the SC SAM, while $\sim44\%$ of the \spr sources are hybrid. In models like the SC SAM the rapid growth of the black hole leads to stopping the star formation activity, while empirical models calibrated against far-IR data like \spr predict a large number of co-evolving systems. \label{fig:galpopsPRIMAger}}
\end{figure*}

State-of-the-art galaxy evolution models show order of magnitude disagreements in their predictions of the distribution of the ratio of BHAR/SFR \citep{Habouzit2020,Habouzit2021}. In general, models calibrated on physical quantities derived from UV/optical/near-IR data (for example, stellar mass, SFR) such as IllustrisTNG and the SC SAM \citep{Somerville2008, Somerville2015} predict faster black hole growth at high masses and luminosities, with the consequent quenching of the star formation activity. Empirical models calibrated on far-IR data from the {\em Herschel} telescope such as \spr or the Simulated Infrared Dusty Extragalactic Sky \citep[SIDES,][]{Bethermin2017}, on the other hand, predict slower black-hole growth and more vigorous dust-enshrouded star formation in luminous systems. The main reason for this discrepancy is that much of the stellar and black hole growth takes place behind large screens of gas and dust that absorbs photons from the near-IR to the soft X-ray bands \citep{Hickox2018}. 

Figure \ref{fig:galpopsPRIMAger} shows a comparison of the $z=1-1.5$ slice of a survey of a galaxy population synthesised using \spr, with that synthesised using the SC SAM over the same 10 square degree light-cone \citep{Yung2022}. These calculations use the depth attainable with the conservative performance of PRIMAger in 150 hours per square degree (i.e., Wide survey in Tab. \ref{tab:depths}), and the extraction depth attainable using a source extraction based on the software XID+ \citep{Donnellan2024} requiring that the sources are detected at $5\sigma$ or better over half the bandpass of PHI. The key difference is that \spr predicts a much larger number of composite AGN-starburst systems, as well as many more galaxies with larger SFRs than the SC SAM. In the SC SAM, only $\sim4\%$ of the galaxies have BHAR$>0.06$~M$_\odot$\,yr$^{-1}$ and SFR$>20$~M$_\odot$\,yr$^{-1}$, while for \spr that fraction is $\sim44\%$. The fairly large discrepancy between models highlights the urgent need for deep surveys with a sensitive, far-IR telescope like PRIMA that can disentangle star formation and AGN powered emission within large numbers of dusty galaxies at cosmic noon. The fast mapping capabilities of PRIMA will enable sample sizes detected with PRIMAger that are $100\times$ to $1,000\times$ larger than those available from \hers \citep{Magnelli2013,Delvecchio2014}, providing us with a true statistical view of black hole and star formation activity during the formative era of galaxies.

\subsection{SFR vs. BHAR with FIRESS using far-IR lines}

\begin{figure}
    \centering
    \includegraphics[trim=20 5 20 20,width=\linewidth,keepaspectratio]{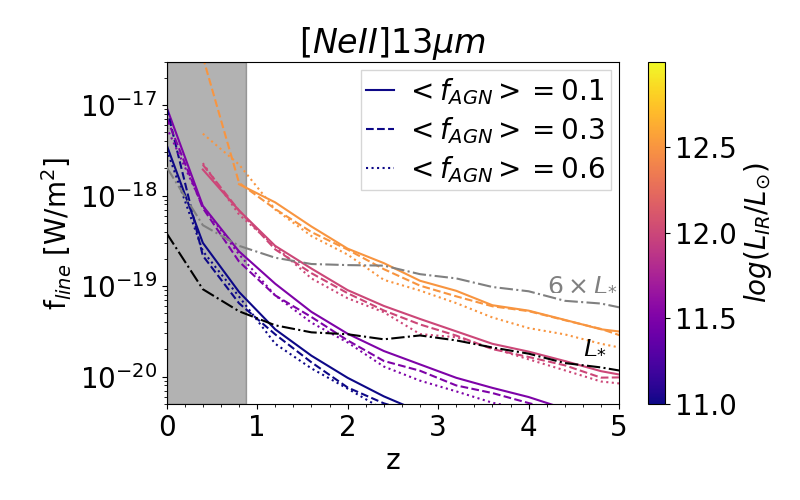}
    \includegraphics[trim=20 30 20 20,width=\linewidth,keepaspectratio]{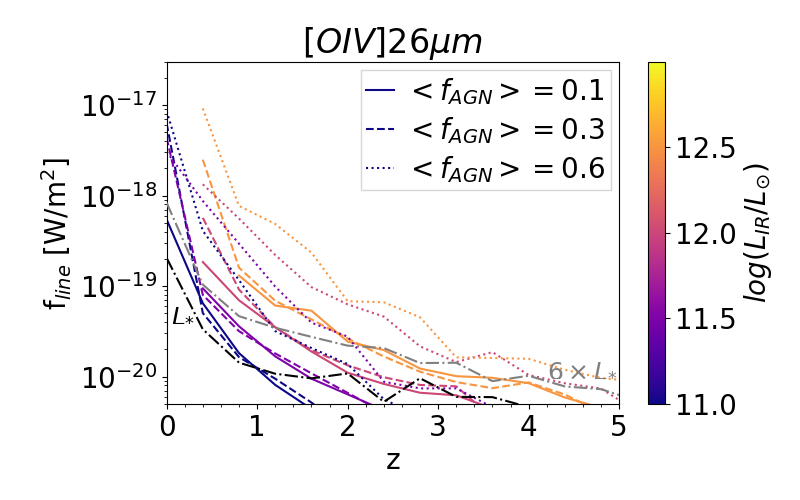}
    \caption{Spectroscopic follow-up of galaxies down to $10^{-19}\,W/m^{2}$ will start probing more normal galaxies and not only ultra-luminous IR galaxies. We show the \neii (top) and \oiv (bottom) line fluxes as a function of redshift, total IR luminosity (see color bar) and IR AGN fraction (different line styles in bins with $\Delta f_{AGN}=0.2$). Dashed dotted lines indicate the line fluxes for galaxies in the knee of the IR luminosity function and six times above it \citep{Traina2024}. The grey area on the top panel shows the redshift range in which the \neii line is not in the PRIMA/FIRESS wavelength coverage.}
    \label{fig:NeII_OIV_fluxes}
\end{figure}

As already mentioned, using far-IR spectroscopy it is possible to have a direct estimate of both the BHAR and the SFR for the same object. Indeed, the \oiv line is a robust calibration of the BHAR, while the \neii line is an accurate indicator of the SFR \citep[e.g.,][]{Spinoglio2015,Stone2022}. In the case of the \oiv line, the contribution excited by star formation can be accounted for \citep{Stone2022}, and the blended contribution from the much fainter nearby \feii line will not affect the measurement. In Fig. \ref{fig:NeII_OIV_fluxes} we show how the predicted flux of these two lines change with the IR luminosity, the AGN fraction (defined between 8 and $1000\,\mu m$) and the redshift, as derived from \spr{}. 

We concentrate on fluxes down to $10^{-19}\,W/m^{2}$, which is feasible for point sources with FIRESS in the low-resolution mode ($R\sim100$) with one hour of integration time. We however highlight that this depth is just chose as an example and it is indeed possible to integrate more time with FIRESS. In details, observations down to $10^{-19}\,W/m^{2}$ allows for tracing the \neii line in galaxies of $L_{IR}=10^{11}\,L_{\odot}$ up to $z=0.7$, while galaxies of $L_{IR}=10^{12.5}\,L_{\odot}$ can be observed up to $z=3$. These fluxes show little dependence on the AGN fraction, given that \neii line mainly traces star-formation. The \oiv line, which efficiently traces AGN activity,  is instead generally fainter than the \neii line, therefore similar observations trace galaxies of $L_{IR}=10^{11.0}\,L_{\odot}$ up to $z=0.4$ for $f_{AGN}<0.5$, but up to $z=0.8$ for higher AGN fractions. At the same time, the \oiv line is brighter than $10^{-19}\,W/m^{2}$ in galaxies of $L_{IR}=10^{12.5}\,L_{\odot}$ up to $z=1$, if $f_{AGN}<0.5$, and up to $z=1.8$ for higher AGN fractions. To put these values in a broader context, we also report the line fluxes for galaxies at the knee of the IR luminosity function, as derived by \citet{Traina2024}, and six times above it . 
The galaxies with \oiv or \neii line brighter than $10^{-19}\,W/m^{2}$ are generally beyond the knee of the luminosity function, but are not tracing hyper-luminous IR galaxies ($L_{IR}>10^{13}\,L_{\odot}$). Moreover, it is necessary to take into account that there will be several other low-ionisation far-IR fine-structure lines that trace star formation or AGN. Correlating these lines could potentially enable us to detect galaxies to greater depth and to give confidence in weaker line detections.

\begin{figure}
    \includegraphics[width=\linewidth,keepaspectratio]{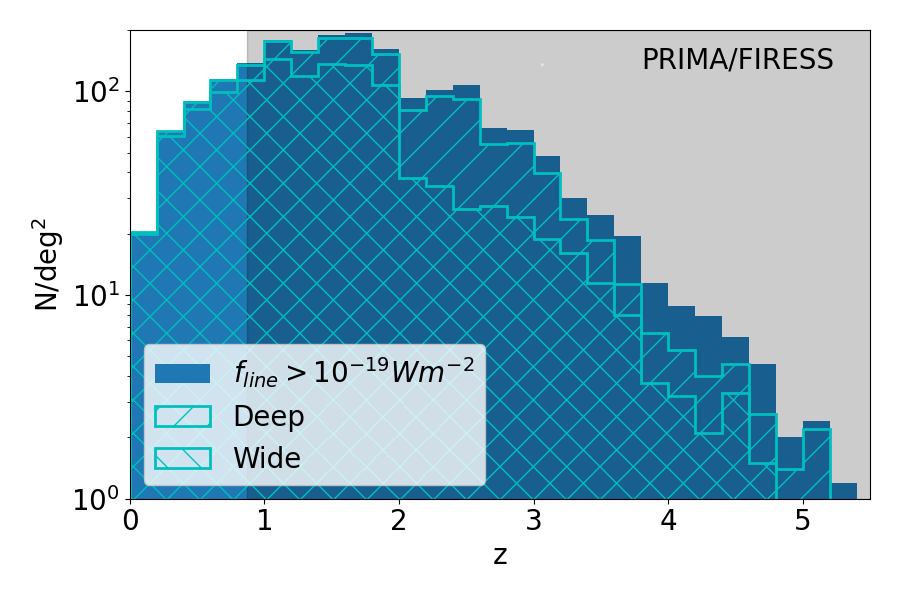}
    \caption{The photometric survey can be used to identify galaxies to follow-up in spectroscopy. We show the redshift distribution of galaxies with \oiv and \neii lines above $10^{-19}\,W/m^{2}$ (blue histogram) with the subsamples detected in the Deep and Wide surveys (hatched histograms), considering the conservative depths. The grey area show the redshift range in which \oiv and \neii lines are in the PRIMA/FIRESS wavelength coverage, but other lines can be used at $z<0.8$ to derive SFR and BHAR.}\label{fig:zsidt_firess}
\end{figure}

In order to simultaneously derive SFR and BHAR  to trace possible signatures of co-evolution between the AGN and its host galaxy, it is necessary to observe both \oiv and \neii in the same galaxies. This will be possible only for a subsample of the aforementioned objects, whose redshift distribution is shown in Fig. \ref{fig:zsidt_firess}. This subsample corresponds to around 19,000 objects/deg$^{2}$ up to $z=5.5$, of which $\sim$15,800 objects/deg$^{2}$ are at $z>0.8$ and inside the FIRESS wavelength coverage. Among the latter, $91\%$ and $59\%$ of the galaxies are expected to be also detected with PRIMAger above the Deep and Wide depths, respectively. In particular, the deep photometric survey includes $>90\%$ and $>50\%$ of objects with \oiv and \neii up to $z=2$ and $z=3.6$. Therefore, the Deep PRIMAger photometric survey can efficiently be used to select target to follow-up in spectroscopy to validate and calibrate the SFR and BHAR derived trought the SED-fitting analysis.

\subsection{The PAH metallicity dependence}

Up to 20\% of the total infrared emission in galaxies emerges as vibrational emission at 3--18$\mu m$ of small carbon-rich PAH grains \citep{Smith2007}.  \textit{Spitzer} observations of PAHs in the local Universe uncovered a strong dependency of the strength of PAH emission on the gas phase metallicity, with an apparent steep decline in the PAH abundance at metallicities of less than $\sim$25\% solar \citep{Engelbracht2005,Smith2007,DraineLi2007,Hunter2010,Sandstrom2012,Aniano2020,Whitcomb2023}. This connection between small grain emission and gas metal abundance ties directly to the lifecycle of PAHs, with indications from changing size and heating-sensitive band ratios of strong evolution in the distribution of grain sizes and potential photo-destruction in high intensity, UV-bright radiation environments.  As explored by \citet[e.g.][]{Calzetti2007}, the metallicity-dependence of PAH emission also impacts their use and calibration as indicators of SFR and AGN fraction.  

\jwst is now uncovering evidence that PAHs are deficient in lower mass (and likely lower metallicity) galaxies approaching cosmic noon \citep{Shivaei2024}, and has also identified unexpectedly strong UV absorption in the 2100$\AA$ feature sometimes attributed to PAHs at redshifts as high as $z=7$ \citep{Witstok2023}.  \jwst spectroscopic mapping observation of a strongly-lensed dusty luminous galaxy at $z=4.1$ revealed strongly varying PAH 3.3$\mu m$ emission \citep{Spilker2023}, in what is currently the earliest known PAH detection.  Yet the majority of PAH power shifts out of \jwst's passbands by $z=2.1$, and the long-wavelength spectroscopic sensitivity of \jwst Mid-Infrared Instrument limits the study of PAH emission to the single 3.3$\mu m$ band which comprises just a few percent of the total PAH luminosity \citep{Lai2020}.  At cosmic noon and earlier times, the bulk of PAH power shifts into PRIMA's far-IR passbands (see Fig. \ref{fig:irlines}).

\begin{figure}
    \centering
    \includegraphics[width=\linewidth,keepaspectratio]{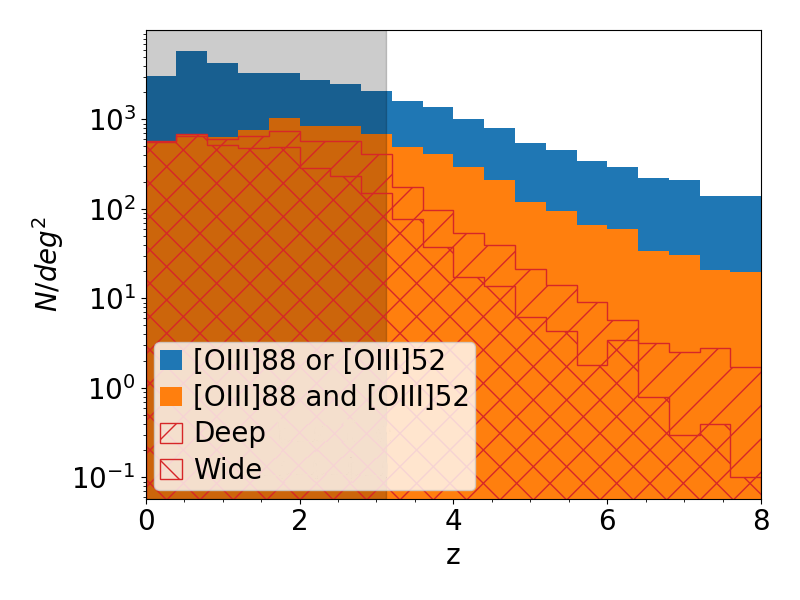}
    \caption{Redshift distribution predicted in \spr for galaxies with \niii, \oiiia and/or \oiiib lines above $10^{-19}\,W/m^{2}$. The hatched histogram shows the sub-population that is also detected in the Deep and Wide PRIMAger surveys, considering the conservative depths. The grey area show the redshift range in which \niii and \oiiia lines are in the PRIMA/FIRESS wavelength coverage.}
    \label{fig:N303_z}
\end{figure}

\spr predicts that samples of the order of $\sim10^4$ PAH-emitting galaxies will be detected in the deep hyper-spectral survey (Fig. \ref{fig:zdistribution}), including luminous sources out to $z=8$.  At z$\sim$2, the metallicity measure based on \niii, \oiiia and \oiiib lines (N3O3 method, Sect. \ref{sec:spritz_lines}) combines with direct hyper-spectral and spectroscopic recovery of PAH luminosity to directly test models of PAH lifecycle and metal sensitivity during the epoch when dust and metal production peaked in the Universe. As visible in Fig. \ref{fig:N303_z}, we expect 850 objects$/deg^{2}$ to have all the three lines necessary for the N3O3 measure (i.e. \niii,\oiiia,\oiiib) above $10^{-19}\,W/m^{2}$ at $z<1.7$, which is the maximum redshift at which FIRESS can observe \oiiib. Almost all ($>98\%$) these galaxies are expected to be detected in the PRIMAger surveys. At $z=1.5-2.5$, we instead expect to observe around 2,000 objects$/deg^{2}$ with both \niii and \oiiia lines above the same flux limit. Moreover, considering the conservative depths, 72$\%$ and 43$\%$ of these objects will be detected in the PRIMAger Deep and Wide surveys, respectively. These fractions change by 1\% or less if we consider the payload sensitives. This shows that the Deep photometric survey can be used as prior to select objects to follow-up with FIRESS. 

\subsection{Predicting and Observing Galactic Outflows}
\label{sec:outflows}

Galactic outflows, whether driven by AGN or the collective effects of supernovae, are believed to play an important role in galaxy evolution. These outflows can heat or even remove entirely the gas available for future star formation as well as reduce or halt gas infall from the CGM. Outflows are known to be very common in luminous and ultraluminous IR galaxies in the local universe \citep[e.g.,][]{Veilleux2013}, with velocities that can exceed 1000~km\,s$^{-1}$. They are invariably multi-phase, with a cool component thought to dominate the mass and momentum of the outflow. This cool component can be observed and characterised directly in the far-IR, most commonly via far-IR OH lines observed in absorption against the far-IR continuum of the host galaxy \citep{Spoon2013,Veilleux2013,Stone2016}. In particular, mass outflow rates can be derived using OH rotational doublets at 65, 71, 79, and 84 $\mu m$ and used in conjunction with radiative-transfer models to estimate the physical parameters of the cool component of the wind, most importantly the mass and mass outflow rates, which can match and in some cases exceed the SFRs even in the most luminous galaxies \citep{Gonzalez-Alfonso2017}.
 
To predict the occurrence of outflows and estimate their properties over cosmic time we need to rely on simulations. Hydro-dynamical simulations, in particular, can be used to model the properties and effects of outflows in individual and ensembles of galaxies, but the detailed physics and implementation varies significantly \citep{Wright2024}. Just as importantly, these simulations are extremely costly to run, and it is infeasible to simulate the large volumes that will be surveyed by future telescopes like PRIMA. All simulations predict strong outflows associated with rapidly growing galaxies and supermassive black holes, although the details vary from simulation to simulation. To estimate the galactic outflows that we can expect to study with PRIMA, we use scaling relations extracted from one such simulation and ``paint'' them on top of the galaxy population in the \spr model. In order to do this, we make use of the galactic outflows in IllustrisTNG \citet{Wright2024}, and statistically relate them to galaxy properties as available in both IllustrisTNG and \spr. We start by converting the BHARs and SFRs to galaxy bolometric luminosities using the standard recipes \citep[assuming an efficiency of 10\% for the accretion, and using the][conversion between SFR and luminosity]{Kennicutt2012}. We then derive a relation between bolometric luminosity and outflow mass loss rate (${\rm \dot{M}_{out}}$), as measured at a scale of $0.1\,r_{\rm vir}$ from the centre of the galaxy halo at redshift $z=2$, where $r_{\rm vir}$ is the halo virial radius. 

\begin{figure}
    \includegraphics[width=\linewidth]{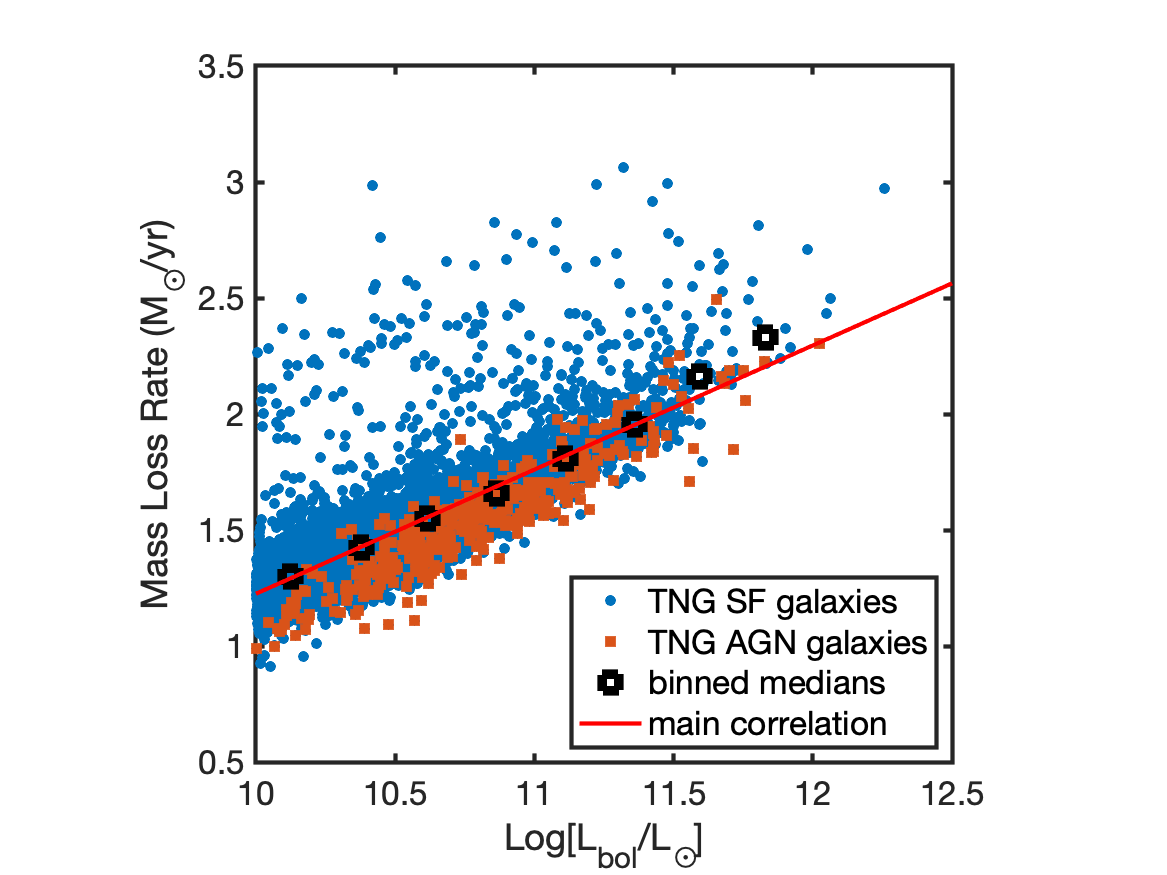}
    \caption{Correlation between bolometric luminosity and outflow mass-loss rate in IllustrisTNG simulations. Red square symbols represent galaxies where L$_{\rm bol}$ has a $\geq20\%$ contribution for the AGN, while blue symbols are all other galaxies. The median mass-loss rate in bins of L$_{\rm bol}$ is shown by the black symbols. The main correlation, $\log({\rm \dot{M}_{outf}})=0.54\,\log({\rm L_{\rm bol}})-4.13$, is shown by the red line. \label{fig:TNGcorrelation}}
\end{figure}

These results in a correlation between the mass outflow rate (in M$_\odot$\,yr$^{-1}$) and the bolometric luminosity (in L$_\odot$), such that $\log({\rm \dot{M}_{out}})=0.54\,\log({\rm L_{\rm bol}})-4.13$, with a tail of objects that scatter toward higher mass-loss rates (Figure \ref{fig:TNGcorrelation}). The relative importance of the tail is larger when considering systems with higher outflow velocities, containing just over 2\% of the systems with outflows at $0.1\,r_{\rm vir}$ to just over 7\% for systems with outflow velocities $\geq250$~km\,s$^{-1}$ at $0.1\,r_{\rm vir}$. Note that a system being in the high-velocity tail of the distribution does not correlate with it harbouring an energetically important AGN at the time of the measurement (red symbols in Figure \ref{fig:TNGcorrelation}), and may instead reflect past AGN activity.

\begin{figure}
    \includegraphics[width=\linewidth]{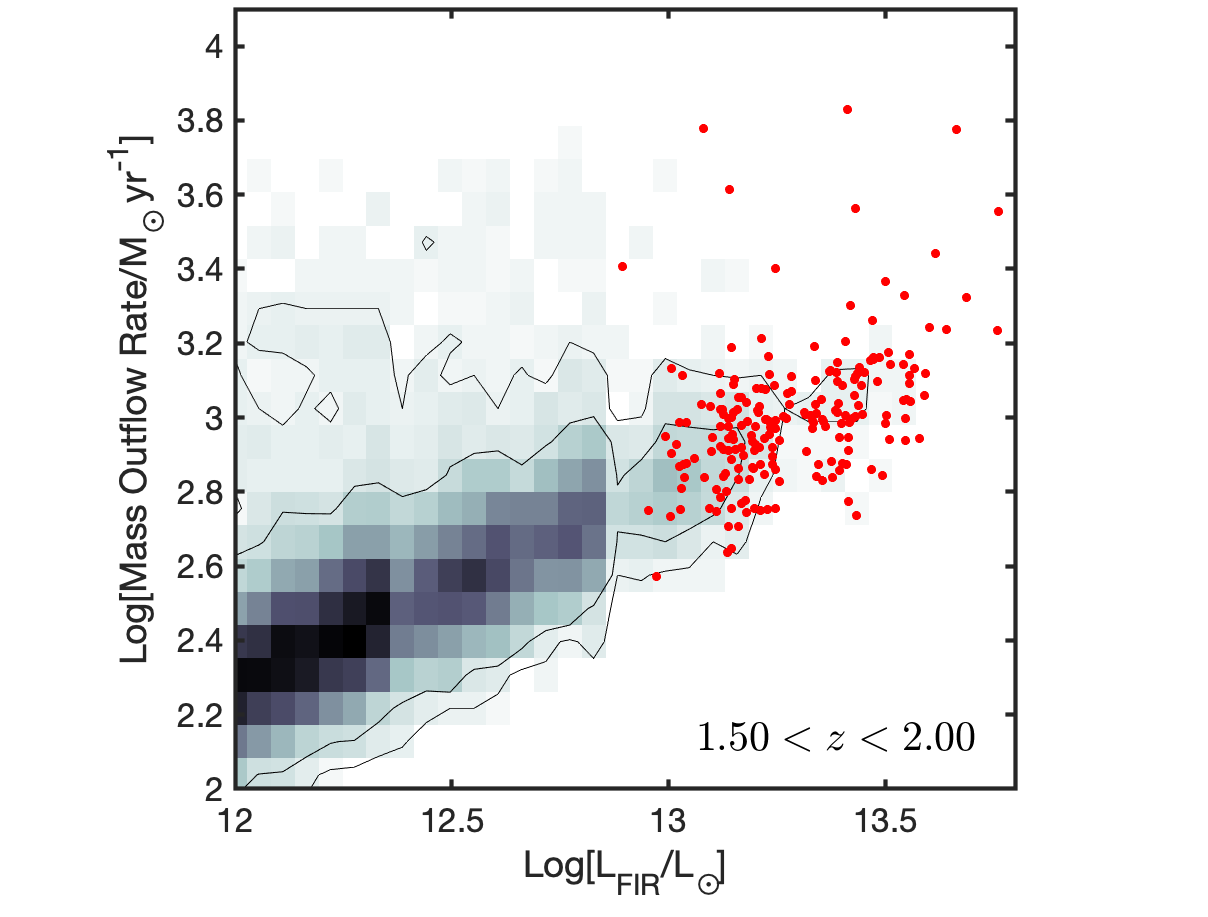}
    \caption{Predicted mass-loss rates in \spr galaxies well-detected with PRIMAger in a 10 sq. deg. survey with 150 hours of integration per square degree, for the redshift range $1.5<z<2$, using an extrapolation of the relation described in \S\ref{sec:outflows}. The red galaxies (about 200 of them) are brighter than 120 mJy at 210 $\mu$m, and thus can be followed up in velocity-resolved OH absorption spectroscopy in a moderate integration. These are very rapidly growing systems, expected to host massive winds. \label{fig:TNGprediction}
    }
\end{figure}

Figure \ref{fig:TNGprediction} shows the result of applying the relation described above to the \spr simulations of the infrared sky. Since the models are sparsely populated at the high luminosity end, an assumption here is that the derived relation can be extrapolated to galaxies with very large bolometric luminosities (approaching or even exceeding $10^{13}$L$_\odot$). This is reasonable since these luminous systems represent dusty galaxies where the stellar mass and/or the central black hole are accreting and growing at their largest rates, and these also are expected to host the most powerful outflows - just as they do at low redshift when they are much more rare. In Figure \ref{fig:TNGprediction} we show the density of systems that would be well detected (those detected at $5\sigma$ or larger over more than 50\% of the PRIMAger Hyper-spectral Imager logarithmic passband) in a PRIMAger survey with 150 hours of integration per square degree, at the conservative performance expected for the instrument. Scaling results obtained for the outflow in the nearby ULIRG Mrk\,231 \citep{Gonzalez-Alfonso2014}, we estimate that it is possible to use FIRESS operating at $R\approx900$ to detect and characterize these winds via the 84$\mu m$ OH line for over 200 luminous IR galaxies detected with PRIMAger at $z\sim1-2$ (those brighter than $120$~mJy at 210~$\mu$m). This will expand, by more than an order of magnitude, the number of sources with powerful, massive outflows studied spectroscopically in the far-IR, and allow us to probe quenching in dusty galaxies at cosmic noon, and the subsequent decline in star formation, for the first time.

\section{Summary and conclusions}\label{sec:summary}
The most active phases of star-formation and black-hole accretion in galaxies are hidden behind large columns of gas and dust. Studying critical phases in the co-evolution of super-massive black holes and galaxies therefore requires the ability to penetrate the resulting extinction, and bring to bear sensitive diagnostic tools to study the physical conditions in distant galaxies. The mid and far-IR part of the spectrum provides abundant well-understood diagnostics of the atomic and molecular gas, as well as the dust, providing a window into star formation, black hole growth, metallicity, and feedback driven galactic outflows in even the most obscured galaxies. In this work, we have demonstrated the photometric and spectroscopic capabilities of the PRIMA mission to address key questions in our understanding of how galaxies and super-massive black holes evolve together over a significant fraction of cosmic time.

In particular, we have shown that multi-tiered photometric surveys conducted with PRIMAger will allow us to detect and study galaxies hundreds of times fainter than were reachable by previous far-IR observatories (e.g. \hers, \textit{Spitzer}), reaching the bulk of the population to cosmic noon and beyond. PRIMAger will be able to observe normal ($10^{11}\,L_{\odot}$), star-forming galaxies up to $z=4$, and it's dense wavelength coverage efficiently probes the presence of PAHs for $z\gtrsim1$ and can also be used to accurately separate AGN and star-formation heating by measuring the shape of the IR dust continuum. Using simulations based on the deepest existing far-IR surveys, along with semi-analytic models of evolving galaxies and dark matter halos, we have made predictions of the galaxy populations that could be observed with PRIMAger and FIRESS, along with quantiative estimates of the accuracies with which we can disentangle and measure the signatures of star formation, black hole growth, gas-phase metallicity, and galactic outflows as a function of IR luminosity, redshift and survey area. In particular, we have verified that, using SED-decomposition we are able to retrieve the AGN fraction with respect to the total IR luminosity with a dispersion of 0.06, the fraction of dust mass contributed by PAH ($q_{PAH}$, measured in percentage) with a dispersion of 0.9, and the total IR luminosity with a scatter of 0.1\,dex (\S\ref{sec:sed-dec}). Because of its ability to map large areas of the sky to great depths, PRIMA can be used to generate and measure large samples of dusty galaxies which are 100$\times$ to 1,000$\times$ larger than those available from \hers \citep{Magnelli2013,Delvecchio2014} out to the epoch of peak star formation and black hole growth. 

Spectroscopic follow up with the FIRESS instrument on PRIMA will reach unprecedented depths in the far-IR, $10^{-19}W/m^{2}$ in one hour of integration at $R=100$, allowing detection of faint fine structure lines and PAH emission features which can be used to directly estimate star formation and black hole accretion rates in distant galaxies. Using FIRESS, it will be possible to measure the \oiv and \neii fine-structure lines, and hence the BHAR and SFR in galaxies of $10^{12.5} L_{\odot}$  up to $z\sim1.8$ and $z\sim3$, respectively. We expect around 16000 object/deg$^{2}$ with both \oiv and \neii lines brighter than $10^{-19}W/m^{2}$ at $z>0.8$, with the large majority ($>90\%$) of them securely detected in at least half of the 12 PRIMAger bands. At the same spectroscopic depths, we expect over 2,000 object/deg$^{2}$ at $z=1.5-2.5$ that could be followed-up with FIRESS to derive metallicity measurements using the \niii and \oiiia emission lines. When combined with measures of the PAH emission features, easily detectable with FIRESS at $z>2$, it will be possible to study the connection between small grain emission and gas metal abundance in thousands of distant galaxies.

Finally, we have used the models and correlations between bolometric luminosity and outflow mass and velocity seen in the local Universe, to predict that with a modest amount of observing time, it will be possible to detect and characterise the galactic outflows with the high-res mode of FIRESS, using the OH molecular absorption feature, in more than 200 galaxies to $z\sim1-2$. These observations will expand, by nearly an order of magnitude, the number of galactic outflows measured in the far-IR and allow us to test whether such outflows are sufficient to quench star-formation in bright IR galaxies over the past 10 Gyr of cosmic time.

\begin{acknowledgements}
The authors acknowledge the scientific help of John Arballo and the PRIMA extragalactic science working groups. LB acknowledges support from INAF under the Large Grant 2022 funding scheme (project "MeerKAT and LOFAR Team up: a Unique Radio Window on Galaxy/AGN co-Evolution").
LC acknowledges support from the french government under the France 2030 investment plan, as part of the Initiative d’Excellence d’Aix-Marseille Université – A*MIDEX AMX-22-RE-AB-101. RJW acknowledges support from the European Research Council via ERC Consolidator Grant KETJU (no. 818930).
\end{acknowledgements}
%
\bibliographystyle{aa} 
\bibliography{main} 
%

\end{document}